\documentclass[12pt, letterpaper]{article}

\usepackage{amsmath,amsfonts,amssymb}
\usepackage{graphicx}
\usepackage{psfrag}
\usepackage{enumerate}
\usepackage{subfigure}

\makeatletter \@addtoreset{equation}{section}

\makeatletter\renewcommand\section{\@startsection {section}{1}{\z@}%
                                   {-3.5ex \@plus -1ex \@minus -.2ex}
                                   {2.3ex \@plus.2ex}%
                                   {\normalfont\large\bfseries}}
\renewcommand\subsection{\@startsection{subsection}{2}{\z@}%
                                     {-3.25ex\@plus -1ex \@minus -.2ex}%
                                     {1.5ex \@plus .2ex}%
                                     {\normalfont\bfseries}}

\usepackage[body={6.5in, 8.5in}, left=1in, top=1in]{geometry}

\newcommand{\be}{\begin{equation}}
\newcommand{\ee}{\end{equation}}
\newcommand{\beq}{\begin{eqnarray}}
\newcommand{\eeq}{\end{eqnarray}}
\newcommand{\fracscr}[2]{\frac{\scriptstyle #1}{\scriptstyle #2}}
\newcommand{\zetao}{\zeta_{0}}

\def\[{\left [}
\def\]{\right ]}
\def\({\left (}
\def\){\right )}

\def\r2{\sqrt{2}}

\def\arcoth{{\rm arcoth}\!}

\def\hi{H_{i}}
\def\hout{H_f}
\def\etals{\eta_{LS}}
\def\rhols{\rho_{LS}}

\def\psils{\psi_{LS}}

\def\x{{\cal X}}
\def\t{\tau}
\def\etav{\eta_v}
\def\c{\ell}
\def\time{t}
\def\Htime{{\cal H}_i t}
\def\H{{\cal H}_i}
\def\vout{V_f}
\def\vi{V_i}

\newcommand{\bbibitem}[1]{\bibitem{#1}\marginpar{#1}}

\newcommand\nn{\nonumber}
\newcommand\eea{\end{eqnarray}}
\newcommand\bea{\begin{eqnarray}}
\newcommand{\sfrac}[2]{{\textstyle\frac{#1}{#2}}}

\def\Label#1{\label{#1}%
  \smash{\hbox to0pt{\raise1ex\hbox{\tiny[#1]}\hss}}}
\def\noLabels{\let\Label=\label}
\def\nobbibitem{\let\bbibitem=\bibitem}

\begin{document}

\begin{titlepage}

\vfil\

\begin{center}

{\Large{\bf  Eternal Inflation, Bubble Collisions, and the\\[.2cm]
Disintegration of the Persistence of Memory }}

\vspace{8mm}

Ben Freivogel\footnote{e-mail: freivogel@berkeley.edu
}, Matthew Kleban\footnote{e-mail: mk161@nyu.edu}, Alberto Nicolis\footnote{e-mail: nicolis@phys.columbia.edu}, and Kris Sigurdson\footnote{e-mail: krs@phas.ubc.ca}
\\

\vspace{7mm}

\vspace{.3cm}
{\small \textit{$^{\rm 1}$ 
Berkeley Center for Theoretical Physics, Department of Physics\\
University of California, Berkeley, CA 94720-7300, USA\\
{\rm and} \\
Lawrence Berkeley National Laboratory, Berkeley, CA 94720-8162, USA
}}

\vspace{.3cm}
{\small \textit{$^{\rm 2}$ 
Center for Cosmology and Particle Physics,\\
Department of Physics, New York University, \\
4 Washington Place, New York, NY 10003, USA }}

\vspace{.3cm}
{\small \textit{$^{\rm 3}$ 
Department of Physics and ISCAP,\\
 Columbia University, 
New York, NY 10027, USA }}

\vspace{.3cm}
{\small \textit{$^{\rm 4}$ 
Department of Physics and Astronomy, \\
University of British Columbia, 
Vancouver, BC  V6T 1Z1, Canada }}


\vfil

\end{center}
\setcounter{footnote}{0}

 
\begin{abstract}
\noindent
We compute the probability distribution for bubble collisions in an 
inflating false vacuum which decays by bubble
nucleation.  Our analysis generalizes previous work of Guth, Garriga,
and Vilenkin to the case of general cosmological evolution inside the
bubble, and takes into account the dynamics of the domain walls that form between the colliding bubbles.  
We find that incorporating these effects  changes the results dramatically:  
the total expected number of bubble
collisions in the past lightcone of a typical observer is $N \sim
\gamma \,  V_f / V_i \, $, where
$\gamma$ is the fastest decay rate of the false vacuum, $V_f$ is its
vacuum energy, and $V_i$ is the vacuum energy during inflation inside
the bubble.  This number can be large in realistic models
without tuning.  
In addition, we calculate the angular position and
size distribution of the collisions on the cosmic microwave background
sky, and demonstrate that the number of bubbles of observable angular
size is $N_{LS} \sim  \sqrt{\Omega_k} N $, where $\Omega_k$ is the curvature contribution to
the total density at the time of observation. The distribution is almost exactly isotropic.

\end{abstract}


\end{titlepage}

\renewcommand{\baselinestretch}{1.05}  


\newpage
\begin{figure}[h!]
\begin{center}
\includegraphics[width=10cm]{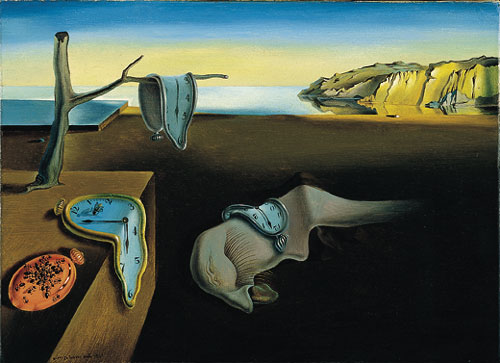}
\end{center}
\caption{\small {\it The Persistence of Memory}, Salvador Dal\'{i}, 1931 }
\end{figure}

\vspace{0.5in}

\begin{figure}[h!]
\begin{center}
\includegraphics[width=10cm]{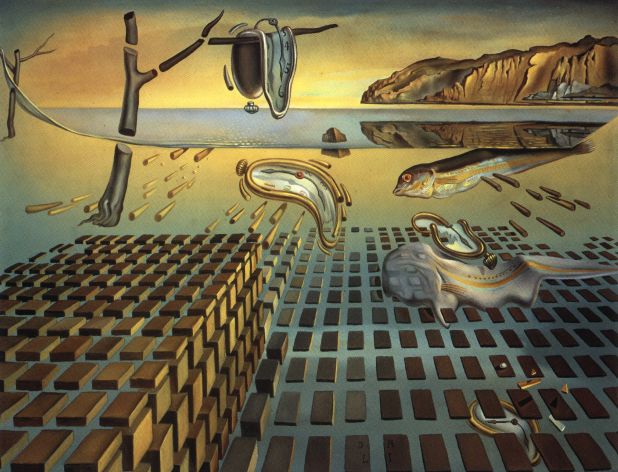}
\end{center}
\caption{\small {\it The Disintegration of the Persistence of Memory}, Salvador
  Dal\'{i}, c.\,1952-1954. 
 }
\end{figure}

 ``My father today is Dr. Heisenberg''
  -Salvador Dal\'{i}, {\it Anti-Matter Manifesto}, 1958

\clearpage

\section{Introduction}\label{sec-intro}

In a model like the string theory landscape \cite{Bousso:2000xa,Susskind:2003kw} one may expect our observable universe to
be contained inside a bubble which is surrounded by an eternally inflating false vacuum.  Quantum tunneling
and thermal effects allow the false vacuum to decay, producing regions
of lower energy---like the one we inhabit---which expand and may
themselves decay in turn.  If the decay rate of the false vacuum in
Hubble units is less than a particular order one number, this process
never terminates. The false vacuum produces new volume through
exponential expansion faster than it loses volume due to decays.

Nevertheless any bubble will collide with many others, forming a
cluster that grows without bound as time passes, as first described in
\cite{Guth:1982pn}.  We would like to ask
whether observers like ourselves can expect to have a bubble collision
in our past. If the answer is yes, then we would like to know the
expected number of collisions, their distribution in size and
direction, and their observable consequences.

This distribution of vacuum bubble collisions was analyzed recently in
a very interesting paper \cite{ggv} by Guth, Garriga, and Vilenkin (GGV)
titled ``Eternal inflation, bubble collisions, and the persistence of
memory.'' 
The analysis of GGV was done in the approximation that the
cosmological constant inside the observation bubble is the same as the
cosmological constant of the false vacuum. Also, GGV assumed that
observers cannot form anywhere in the future lightcone of collisions.

With these assumptions, GGV computed the probability distribution for
bubble collisions which may become visible in the
future. 
They found that a typical observer
inside such a bubble sees an anisotropic distribution of
collisions. The anisotropy is left over from the initial conditions,
and does not go away even in the limit that the observation bubble
nucleates an infinitely long time after the initial conditions.
They called this lingering effect of the initial conditions ``the persistence of memory."

However they also found that the overall rate at which an observer sees
collisions (and the probability of observing any collisions) is
proportional to the dimensionless decay rate $\gamma$ of the false
vacuum, a quantity that is typically exponentially small.  Moreover
they found that the typical distance to the nearest collision is
large, of order $- \ln \gamma$ in units of curvature radius.
 
 In this
note we consider the case relevant for observation, where the interior of
the observation bubble is similar to our
universe. We assume that due to the dynamics of the domain walls, 
collisions disrupt structure formation
in only part of their future lightcone (as described in more detail in
section \ref{dwsec}).
We find that these changes have a major impact on the result. The
ratio of the vacuum energy in the false vacuum to the inflationary
vacuum energy,
$\vout / \vi$, which was set to one in the GGV approximation, can be large in
realistic scenarios. Among other things, this ratio turns out to divide the
anisotropic term in the distribution and multiply the expected number
of bubbles in the past lightcone of a late-time observer.  Incorporating
the effects of domain walls renders the distribution smooth and finite
even for observers at infinite ``boost".

A heuristic argument for the distribution of bubbles in our past lightcone
is as follows. 
In the inflating
false vacuum there is an event horizon with radius of $R_f = H_f^{-1}$
(where $H_f\sim\sqrt{\Lambda_f}$ is the false vacuum Hubble constant)
surrounding every point.  This horizon shields the bubble from
interacting with any event (such as another bubble's nucleation) that
occurs more than a distance $H_f^{-1}$ away from its walls.  Hence the
bubble wall is surrounded at all times by a shell $H_f^{-1}$ thick
inside of which any bubble nucleation leads to a collision.  As the
bubble grows the volume of this region expands like its surface area,
and at the same time the relative size of the late-appearing bubbles
shrinks.  This leads to a size distribution of bubble collisions
which, as a function of angular radius on the fiducial bubble,
increases rapidly with decreasing size. Further, as bubble grows the
initial conditions become unimportant, so the anisotropy
is insignificant for small bubbles.

To compute the total number of bubbles, we need to know the 4-volume in
the false vacuum available to nucleate bubbles that will collide with
our own. The
surface area of the bubble wall inside the past lightcone of
late-time observers like us is of order $H_i^{-2}$, where $H_i$ is the Hubble
constant during slow-roll inflation inside the bubble. As mentioned
above, the distance from the bubble wall to the event horizon is
$\hout^{-1}$. To get the 4-volume, we just need to know the amount of
time available for creating collision bubbles, which turns out to also be
given by $\hout^{-1}$.
 
As a result
the 4-volume of the false vacuum available for bubble nucleations
scales as $H_f^{-2} H_i^{-2}$.  If $\Gamma$ is the decay rate per
4-volume, the total number of bubbles in our past lightcone is $
N \sim \Gamma \hi^{-2} \hout^{-2}$
Defining the dimensionless decay rate $\gamma \equiv \Gamma
\hout^{-4}$, we have
\be
N \sim  \gamma \left(H_f \over
  H_i \right)^2~.  
\label{roughn}
\ee

It is crucial in deriving this answer that observers can form in a
reasonable fraction of the future lightcone of the collision. Whether
this is true depends primarily on the dynamics of the domain wall
that is formed in the collision. The domain wall will have constant
proper acceleration and approach the speed of light. Depending on parameters it may accelerate
into our bubble or away from our bubble. 

If the domain wall accelerates into our bubble, then most of the
future lightcone of a collision is uninhabitable because it is eaten
by the collision bubble, and the probability of observing a collision
will be small, as in GGV. Further, such an observation would be
followed nearly instantaneously by a fatal collision with the domain
wall.

We choose to analyze the more optimistic case where the domain wall
formed in the collision accelerates away from our bubble. As we
describe in section \ref{dwsec}, it is reasonable to expect that in
this case an order one fraction of the future lightcone of the
collision is inhabitable, and the estimate (\ref{roughn})
holds. 

Finally, one could also analyze the case where our bubble
collides with identical bubbles. In this case no domain wall forms
after the collision, and again we expect an order one fraction of its
future lightcone to be inhabitable. We will focus on the case where
the collision produces a domain wall, but we expect that our analysis
carries over almost unchanged to the collision of our bubble with
identical bubbles.

\subsection{Results}

Our primary result is that the expected number of bubble collisions
in our past is 
\be N \approx {4 \pi \over 3} \gamma \left(H_f \over
  H_i \right)^2~.  
\ee 
Here $\gamma \sim e^{-S}$ is the dimensionless
decay rate of the false vacuum to its fastest decay product.  The
inflationary Hubble scale $H_i$ is very poorly constrained, but
assuming the vacuum energy in the false vacuum is of order the Planck
scale, observational bounds on tensor modes imply that $ (H_f/H_i)^2 \gtrsim
10^{12}$, and it can be larger than $10^{80}$ in models of low scale
inflation.  Also, the decay rate is bounded below by the recurrence
time of the false vacuum de Sitter space, $\gamma > \exp(-S_f)$ with
$S_f \sim \left( M_P / \hout \right)^2$. If the vacuum energy of the
false vacuum is near the Planck scale the decay rate cannot be too
small. So even though bubble nucleation is exponentially suppressed,
the total number of bubbles in the past lightcone of a typical
observer can be large with no fine-tuning.

  We derive the distribution of bubble collisions in the backward
  lightcone of a single observer as a function of the angular
  coordinates $\theta$ and $\phi$ of the center of the collision
  bubble, a parameter $\x$ that controls the center of mass energy of
  the collision, and the conformal time $\eta_v$ when
  the collision first becomes visible ({\bf Eq. \ref{dist}}).

  We find that the ``persistence of memory" effect discovered by GGV
  is significantly modified when the vacuum energy during inflation is
  allowed to be different from the cosmological constant of the false
  vacuum.  The memory of the initial conditions, which manifests
  itself as an anisotropy in the angular distribution, disappears in
  the limit that the bubble has small cosmological constant
  inside ({\bf Eqs. \ref{dist} and \ref{kdist}}) and the
  late-time distribution becomes precisely boost and rotationally
  invariant around the bubble's nucleation point ({\bf
    Eq. \ref{Bdist}}).

Finally, we derive the distribution of collision lightcones on the last scattering surface as a function of their position and angular size on the CMB sky ({\bf Eq. \ref{psils}}).  Bubbles that collide with ours very early have lightcones that completely cover the part of the last scattering surface we can see, and these are probably not observable in the limit that the size of the region they cover is much greater than the observable part.  But bubble collisions that occurred somewhat later have lightcones that bisect the observable part of the last scattering surface (and hence the CMB sky), and these can directly affect the cosmic microwave background temperature map and are of observational interest.  The total number of such bubbles is 
\be
N_{LS} \approx 4 \zetao N \sqrt{\Omega_k} =  {16 \pi  \over 3} \zetao \gamma \left(H_f \over H_i \right)^2 \sqrt{\Omega_k}~.
\ee
where $\zetao \approx 10/3$ is a constant the depends upon the cosmological evolution inside the bubble after reheating (see Appendix~\ref{app:conf} for details)  and $\Omega_k$ is the curvature contribution to $\Omega_{\rm tot}$.

\subsection{Related work}
\label{related}
Our work is closely related to the analysis of Aguirre, Johnson, and
collaborators \cite{aj1, aj2, aj3}, with the crucial difference that
we treat the effects of bubble collisions on observer formation in a different (and in our opinion more realistic) way.   Our results are in rough agreement with \cite{aj1}
for the number and distribution of ``small'' bubbles, where ``small'' here means
bubbles that are smaller than the observation bubble in the
observer's reference frame. The number of such bubbles was computed in
\cite{aj1} and their isotropy was noted by \cite{aj1} and \cite{bfy}.
However because of our assumptions about the disruptive effects of
collisions on observer formation, our results for the distribution of
large bubbles and our conclusions for the potential observability of collisions differ sharply from \cite{aj1, aj2, aj3}.

Ref. \cite{aj1}
focused on the case where bubble collisions have approximately no
effect on observer formation, and found that typical observers see an
arbitrarily large number of very large, anisotropically distributed
bubbles.   The infinity is removed here by  our treatment of the domain walls. 
Ref.~\cite{aj3} did a numerical analysis of a single collision and
found that the region far to the future of the resulting domain wall
has an approximate $SO(3,1)$ invariance restored (similar to the case
of a single bubble without a collision), which may remain if no other decays or collisions
take place to perturb it.
Based on this and on the conclusions of GGV, \cite{aj3}
concluded that the fraction of observers that can see a collision is
very small.  However, even under this set of assumptions computing this fraction unambiguously is difficult, as it requires comparing two  infinite volume regions.

After this work was completed and while the manuscript was in
preparation we received Ref.~\cite{dahlen}, which
concluded that the fraction of observers that see collisions can be ${\cal O}(1)$, but performed the
analysis in the approximation that the domain walls between bubbles do
not accelerate and that the observation and collision bubbles are
identical and dominated by vacuum energy inside.  In this work we will
take the acceleration of the domain walls into account and allow for a
fully realistic cosmology inside the observation bubble.

The motion of the post-collision domain walls was 
studied in \cite{hms} and \cite{fhs} in special cases, and then more generally in \cite{ckl1} 
and {\cite{aj2}.

Determining the observational consequences for observers in the space to the future of a bubble collision is a rich problem that has been addressed by \cite{aj1, aj2, aj3} and by  \cite{ckl1, ckl2}.   In particular \cite{ckl2} studied in detail the effects on the cosmic microwave background of collision lightcones that bisect the observable part of the last scattering surface.

\section{Probability distribution for bubble collisions}
We want to compute the probability distribution for bubbles that
collide with our own. Before including other bubbles, the exterior of
our bubble is in the false vacuum. We work in the approximation that the
domain wall of our bubble is lightlike, and that the metric outside
the lightcone is undisturbed de Sitter space with cosmological
constant $\hout^2$. This approximation does not have a large
effect on our results, and going away from it would introduce new
model-dependent parameters.

A convenient coordinate system that covers the region where collision
bubbles can  nucleate is
\be
ds^2 = \hout^{-2}{1 \over \cosh^2 \x} (d \x^2 +  d S_3^2)
\ee
where $d S_3^2$ is the metric of a unit $2+1$ dimensional de Sitter
space and $\hout$ is the Hubble constant of the false vacuum. 
Boosts that leave the observation bubble fixed act as the full
symmetry group of the $2+1$ de Sitter space, while leaving $\x$
fixed. Therefore collisions with different $\x$ are physically
different, while any two collisions with the same $\x$ can be boosted
into each other by acting with the symmetries. 

Choosing coordinates on the de Sitter space, the metric is
\be
ds^2 = \hout^{-2} {1 \over \cosh^2 \x} (d \x^2 - d\t^2 + \cosh^2\t
d\Omega_2^2)
\label{outcoords}
\ee
where $\x$ and $\t$ run from $-\infty$ to $\infty$. These coordinates
cover the region outside our bubble that is capable of nucleating
bubbles that will collide with our own. The nucleation point of our
bubble is at $\x = - \infty$ and $\t$ finite, while the domain wall
of our bubble is the null surface $\x = - \infty$ with $\t \geq
0$. The null surface $\x = \infty$ bounds the region where nucleated
bubbles will collide with our bubble.

We are interested in observing bubble collisions from inside the
bubble, while the coordinate system above covers the outside of the
bubble. 
For the moment we leave the cosmological history general inside the
bubble. Before considering the effects of collisions, the interior
geometry has $SO(3,1)$ symmetry. The metric can be written
\be
\label{metric}
ds^2 = a^2(\eta)(-d\eta^2 + d\rho^2 + \sinh^2 \rho d \Omega_2^2),
\ee
where without loss of generality we focus on an observer at $\rho = 0$.  This metric describes
a homogeneous and isotropic Robertson-Walker cosmology.  An unperturbed observation bubble
will inflate and reheat along slices of constant $\eta$, and comoving observers move on trajectories
with fixed spatial coordinates\footnote{Note that we are using conventions
where the scale factor $a$ has units of length, and the spatial
curvature is $k = -1$.}.

This metric is smooth all the way down to the ``big bang" $\eta =
-\infty$, even though $a(\eta)\rightarrow 0$ there.  The reason for
this non-singular ``big bang" is that the instanton boundary
conditions require \cite{fkrs}
\be
a(\eta) \approx  {2 \over \hout} e^{\eta}  + {\cal O}(e^{3 \eta}) {\rm
  \ \
 as \ } \eta \rightarrow -\infty
\ee
where the multiplicative factor $2/H_f$ is arbitrary and has been chosen
for convenience.  For example if one follows the past lightcone $\rho
+ \eta = \eta_0$ of an observer at time $\eta_0$, the radius of the
bubble lightcone is $a(\eta) \sinh(\rho) \rightarrow e^{\eta + \rho}/
H_f = e^{\eta_0}/ H_f$, which is finite as $\eta \rightarrow -\infty$
and $\rho \rightarrow \infty$.

Consider a bubble that nucleates at a point $(\x, \t, \theta,
\phi)$. 
The collision occurs along a spacelike hyperboloid
that preserves $SO(2,1)$ symmetry, as shown in figure \ref{conefig}.
\begin{figure}[h!]
\begin{center}
\subfigure
{\includegraphics[width=6cm]{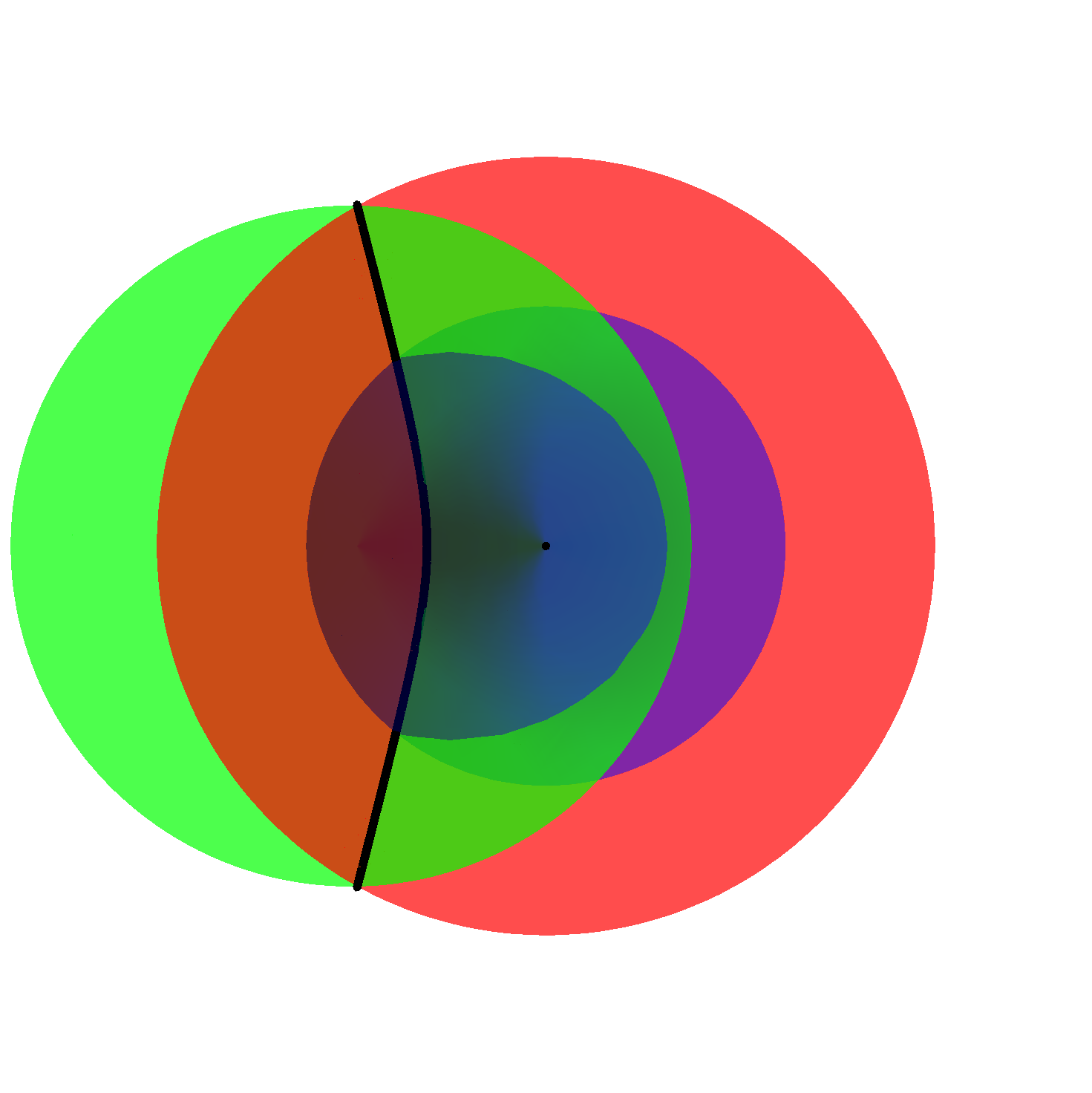}}
\subfigure
{\includegraphics[width=10cm]{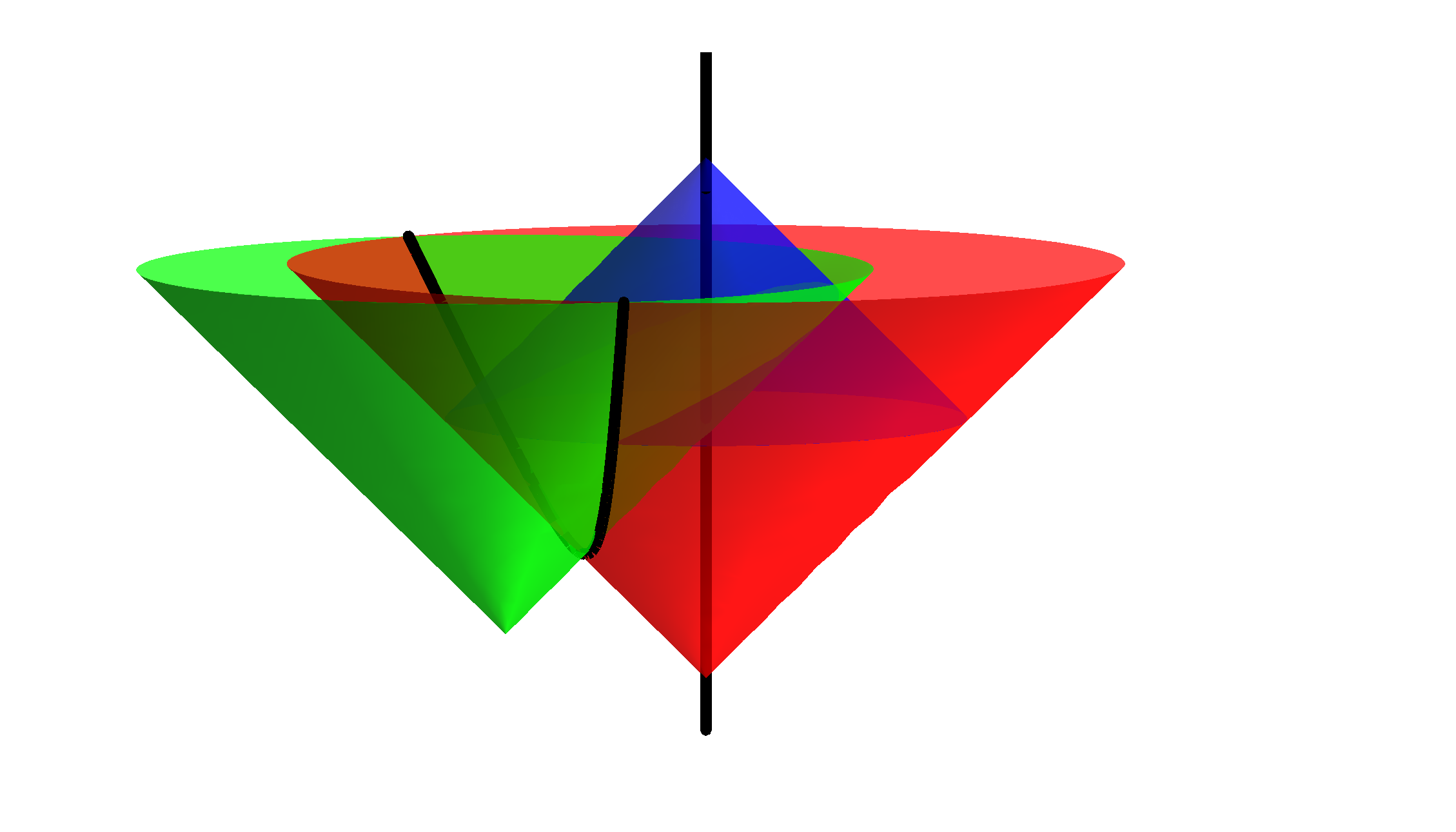}}
\caption{\small Two views of the collision of two bubbles in spacetime.  The red cone is the observation bubble lightcone, the green cone is the collision bubble lightcone, the straight black line is the worldline of the observer, and the hyperbolic black line is the surface along which the bubbles collide.  The angular size of the collision as seen by the observer at some time is determined by the intersection of the observer's past lightcone (in blue) with the observation bubble wall, or with the surface of last scattering. }
\label{conefig}
\end{center}
\end{figure}
The bubble will appear as some type of disturbance on the
observer's sky. The angular location of the center of the disturbance
is given by the angular location $(\theta, \phi)$ of the bubble
nucleation. 
We would like to compute the time at which the collision
first enters the backward lightcone of the observer at
$\rho=0$. Ingoing radial null rays satisfy $\x + \t = c$ outside the
bubble and (with our choice of coordinates) $\eta + \rho = c$ inside the bubble, where $c$ is an
arbitrary constant. So the future lightcone of a bubble nucleated at $(\x, \t)$ 
first reaches $\rho = 0$ at the conformal time
\be
\etav = \x + \t.
\label{etaveq}
\ee

A given collision is described by four parameters: two angular coordinates $(\theta, \phi)$ give the angular location of the center of the collision, $\etav$ is the time when the collision first becomes visible, and $\x$ sets the intrinsic properties of the collision. For example, the radius of curvature
of the collision hyperboloid is $r_c = \exp(\x )$.

The probability to nucleate a bubble in an infinitesimal region is proportional
to the 4-volume of that region,
\be
dN = \gamma \hout^{4} dV_4,
\ee
where $\gamma$ is the dimensionless nucleation rate. The infinitesimal 4-volume in
our coordinates (\ref{outcoords})
is 
\be\label{vol}
dV_4 =  \hout^{-4} {d \x \over \cosh^4 \x} \cosh^2 \tau d\tau d^2 \Omega_2.
\ee
We can exchange the coordinate $\t$ for the conformal time $\etav$ at which the collision enters the backward lightcone by using (\ref{etaveq}). So the naive probability distribution for collisions is
\be
dN = \gamma {\cosh^2(\etav - \x) \over \cosh^4 \x} d \etav d\t d^2 \Omega_2
\ee
Again naively, the total number of collisions visible to an
observer at conformal time $\eta_0$ is given by integrating the above equation over the entire range of parameters, subject to the constraint $\eta_v < \eta_0$.  However this integral diverges at $\etav \to - \infty$, corresponding
to an infinite probability to have nucleated a bubble in the
arbitrarily far past. 

There are several reasons the above analysis is naive, which is fortunate since the answer is infinite. We  discuss various corrections throughout the rest of the paper. 

\subsection{Disruptive effects of collisions}
\label{dwsec}
The most important correction is that bubble collisions will disrupt structure formation in some part of their future lightcone. Since we are interested in making a prediction for our observations, we need to compute the distribution of bubble collisions {\it consistent with the presence of observers.} We show in this subsection that this requirement regulates the infinity of the naive analysis and allows for a well-defined prediction for the distribution of bubbles..
 
The collision can disrupt
structure formation throughout its future lightcone, but it is
reasonable to expect that in some regions the bubble collision has a
small effect, while in other regions no structure forms.
 We focus on the
collision of our bubble with some other type of bubble, so that a
domain wall forms in the collision. Vacuum energies and domain wall
tensions are generically Planckian in the landscape, so we assume that
both the false vacuum and the collision bubble have Planckian vacuum
energy, and that the domain walls in the problem also have Planckian
tension. The vacuum energy in the collision bubble could be
negative or positive.

We make the approximation that structure formation occurs unchanged in
the observation bubble, except that the domain wall with the collision
bubble removes regions that otherwise would have formed
structure. Because the vacuum energy in the collision bubble is
Planckian, no observers can form in the collision bubble.

 The key question is regarding the dynamics of the domain
wall. Because we have assumed that the vacuum energy of the collision
bubble
 and the tension
of the domain wall are Planckian, the acceleration of the domain wall
is generically Planckian. But the domain wall may accelerate into the
observation bubble, or away from the observation bubble. In either
case the domain wall undergoes constant proper acceleration and
asymptotically appraches the speed of light. 

If the domain
wall accelerates into the observation bubble, then nearly all of the
future lightcone of the collision is behind the domain wall, and since
we have assumed that observers can only form in the observation bubble
there are essentially zero observers in the future lightcone of the collision.

If the domain wall accelerates away from the observation bubble, then
some observers will form in the future lightcone of the collision. We
focus on this case. If the vacuum energy of the collision bubble is
positive, then it is guaranteed that the domain wall accelerates away
from the observation bubble. If the vacuum energy of the collision
bubble is negative, then the acceleration of the domain wall depends
on the tension; for sufficiently large tension the domain wall again
accelerates away from the observation bubble.

The detailed dynamics of the domain wall are somewhat involved and
model-dependent \cite{hms, fhs,ckl1, aj2}, but with our assumptions Planckian acceleration
is generic. The initial velocity of the domain wall after the
collision is also model dependent, but typically it will take some
time for the messy collision to settle down into a coherent domain
wall; with our assumptions this time scale is also
Planckian. In order to have some control over our
calculation we assume that the energy scales are all somewhat less than
Planckian, and instead set by the outside Hubble constant $\hout$.\footnote{
In fact, the typical energy scale associated with the vacuum energy, the tension of the wall, and its acceleration, is the geometric mean between the Planck scale and $H_f$. In the following we will ignore this point, for it does not lead to any substantial modification of our results.}

 Therefore, we adopt the following caricature of the domain
wall motion:  we assume the domain wall moves into the observation
bubble at the speed of light for a time $\hout^{-1}$. Then it
turns around and moves outward at the speed of light. While this
approximation could certainly be improved upon in specific models, it
captures the crucial physics of the problem.

As discussed in the paragraph below Eq. \ref{metric}, the metric has a coordinate singularity $a \to 0$ as
$\eta \to -\infty$.  With our choice of coordinates $\eta=0$ corresponds to a time of order
$H_f^{-1}$.
Therefore we will approximate the domain wall trajectory as null
and ingoing for $\eta<0$, null and outgoing for $\eta>0$.  Bubbles whose domain walls cover $\rho=0$ at any time during their evolution are assumed to disrupt
reheating. Therefore, given our assumptions, we prohibit
bubbles with $\etav < 0$.

\subsection{Summary}
Now we can compute the distribution of bubbles that are in the
backward lightcone of an observer at the conformal time $\eta_0$.  It is
\begin{eqnarray} \label{dist}
dN &=& \gamma  { \cosh^2 (\eta_v - \x) \over \cosh^4 \x} d\x d\eta_v d^2 \Omega_2 \\
0 &<& \eta_v < \eta_0
\end{eqnarray}
where the restriction $0 < \etav $ comes from requiring the observer formation is not disrupted.

We can integrate out $\x$ to get the distribution as a function of
conformal time,
\be
dN = {2 \gamma \over 3}(1 + 2 \cosh 2 \eta_v) d\eta_v d^2 \Omega_2 \ \
\ \ \ \ \ \ \ {\rm for }\ \eta_v > 0
\ee

Integrating, the total number of collisions visible before some conformal time
$\eta_0$ is
\be
N(\eta_0) = {8 \pi \gamma \over 3} \left( \sinh 2 \eta_0 + \eta_0
\right)~.
\label{numberofc}
\ee
There is an important subtle point that we have glossed over here. The divergence present in the naive distribution means that the probability of observer formation at $\rho = 0$ is zero. Therefore, one could ask whether the above distribution is meaningful. The answer is that  any choice of initial conditions will regulate the infinity so that the probability of observer formation is finite. Further, the actual choice of initial conditions will have little effect on the distribution, so that the net effect is simply that we can take the above equations seriously. 

\section{Effect of the initial condition surface: how persistent is memory?}

Because the false vacuum is only metastable, it is not consistent to imagine that it exists eternally.
To have a well-defined Cauchy problem, one needs to choose an initial condition on a spacelike or null
surface.  An obvious (but ad hoc) choice is to put the fields in a
false-vacuum minimum on some slice of constant de Sitter time and then
consider the forward time evolution.  One of the surprising results of
the work of \cite{ggv} was that observers inside bubbles that form to
the future of this initial condition surface always retain some memory
of it no matter how far in the future they live, in the sense
that the angular distribution of incoming bubbles remains
significantly anisotropic 
except for the special single observer in the case
of an isotropic initial condition surface who sees an isotropic distribution.  We will see that
incorporating a realistic cosmology inside the bubble in the analysis
alters this conclusion significantly.

For an initial condition we follow \cite{ggv} in taking the surface to
be $t = - \infty$ in the standard de Sitter flat slicing.  This choice
has the advantage that it is infinitely far back in the past, but
still regulates the divergence in Eq. (\ref{vol}).  On the other hand,
\cite{ggv} found that with this choice only a single special observer
co-moving inside the observation bubble will see an isotropic
distribution, because generic co-moving observers are boosted with
respect to the flat slicing.

For this discussion, since we are interested in comparing different observers related by boosts around the nucleation point of the bubble, it will be
helpful to use the embedding coordinates for de Sitter space in
which all of the symmetries are manifest. $3+1$ dimensional de Sitter
space can be embedded in $4+1$ dimensional Minkowski space as the
surface
\be
X^2 + R^2 - T^2 = 1
\ee
where $R^2 = Y^2 + Z^2 + W^2$ and we have set
\be
\hout = 1 \ {\rm in\ this\ section.}
\ee
The initial condition surface in embedding coordinates is
 $X + T = 0$.

 The relation to the parameters of the bubble collision is
\begin{eqnarray*}
T &=& {\sinh \t \over \cosh \x} \\
X &=& - \tanh \x \\
Y &=& {\cosh \t \over \cosh \x} \cos \theta \\
Z &=&  {\cosh \t \over \cosh \x} \sin \theta \cos \phi \\
W &=&   {\cosh \t \over \cosh \x} \sin \theta \sin \phi
\end{eqnarray*}
where we have chosen the observation bubble to nucleate at $X = 1, T = Y = Z = W = 0$.
Recall that the coordinate $\t$ is related to our preferred parameter $\etav$ by $\etav = \t + \x$.

 With this choice for the initial condition surface, allowed collision bubbles nucleate in the region
$X + T > 0$.  Translating to the parameters of the bubble collision,
this condition is 
\be
- \tanh \x + {\sinh \t \over \cosh \x} > 0~,
\ee
which is equivalent to
\be
\x < \t ~.
\ee
This is the appropriate condition for the special ``unboosted"
observer (the observer that is comoving in the bubble and at rest in the flat slicing). To consider boosted observers, we can simply boost the initial
condition surface. The boost should preserve the nucleation point of the
observation bubble; consider without loss of generality a boost in the $Y T$ plane. The boosted
initial condition surface is
\be
X + T \cosh \xi + Y \sinh \xi > 0
\ee
where $\xi $ parameterizes the boost. In our coordinates this
condition becomes
\be
\sinh \x < \sinh \t \cosh \xi + \cosh \t \sinh \xi \cos \theta~.
\label{eqn:4volrestriction}
\ee
The 4-volume available for bubble nucleations is restricted
by the above inequality. 

Now the initial condition surface is ambiguous, because we
do not know our boost $\xi$. However, we will show that the
  expected number of bubbles that satisfy the other constraints but
  are prohibited by the initial condition surface is generally much less than
  one! Therefore, the unknown parameter $\xi$ has a small effect on
the final distribution.

To show this, we will compute the total number of collisions in the past lightcone
of an observer in a cosmology like ours that are allowed by the other constraints but
prohibited by the initial
condition surface.
The answer for arbitrary $\xi$ is easy to calculate but is slightly complex, so we will just
present the answer at zero and infinite boost. 

At zero boost
the initial condition surface imposes $\x < \t$, so the expected number of
bubbles that are allowed by the other constraints but prohibited by
the initial conditions surface is
\be
\Delta N(\xi = 0) = \gamma \int d \x \int d \t \int d^2\Omega {\cosh^2 \t \over
  \cosh^4 \x}
\ee
integrated over the region
\be\label{range}
0 < \x + \t < \eta_0 \ \ \ \ \ \x > \t
\ee
Evaluating this integral gives $ \Delta N(\xi = 0) = (4 \pi \gamma /3) (1 + \ln 4) +  {\cal O}(\gamma e^{-\eta_0})$.

At infinite boost, the initial condition surface requires
$
\cos \theta + \tanh \t > 0.
$
Performing the integral, the number of bubbles prohibited by the
initial condition surface only is $\Delta N(\xi = \infty)=  (4 \pi \gamma/3)(\eta_0 + 1) +  {\cal O}(\gamma e^{-\eta_0})$.

We will show in the next section that the conformal time $\eta_0$ for an observer like us is
approximately
\be
\eta_0 \approx \log {\hout \over \hi}
\ee
where $\hi$ is the Hubble constant during slow roll inflation.  Since one expects $\gamma$
to be exponentially small, generically $\Delta N(\xi) \ll 1 ~~ \forall ~\xi$.  

\section{Observability of Collisions}

In this section we will use this result to determine the total number of collisions that are 
potentially detectable by an observer in a realistic cosmology.

\subsection{Computation of the conformal time}
\label{sec:conf}

To justify our claim that $\eta_0 \approx \log {\hout \over \hi}$, we need to relate $\eta$ to the proper time
of a comoving observer in the observation bubble.
The conformal time $\eta$ is related to the observer's proper time $t$
by $d \eta = dt/a(t)$. The constant of integration is fixed by
demanding that
\be
a(\eta) \to {1 \over 2 \hout} e^\eta \ \ \ {\rm as} \  \eta \to
-\infty
\label{conv}
\ee
 as mentioned in the discussion below Eq. (\ref{metric}).

 To proceed we will need an explicit form for $a(t)$. Approximating
 slow-roll inflation by de
 Sitter space with Hubble constant $H_i$, $a(t) = H_i^{-1} \sinh(H_i
 t)$ for $t<t_e$. During this time, the scale factor as a function of
 conformal time is
\be
a^2(\eta) = {\hi^{-2} \over \sinh^2(\eta - \log{\hout \over \hi})}
\ee
where the additive constant in $\eta$ is fixed by the convention
(\ref{conv}).

Defining the number $N$ of efoldings by $a_{RH} \equiv \hi^{-1}
e^N$, the conformal time at reheating is
\be
\eta_{RH} = \log{\hout \over \hi} - e^{-N} \approx \log{\hout \over \hi}
\ee
for $a_{RH} \gg \hi^{-1}$.

After reheating, the behavior of the scale factor depends upon which type of matter or energy is dominating the expansion of the Universe at a particular epoch. It is approximately
  $a(t) = H_i^{-1} e^N
 (t/t_{RH})^p$, where $p=1/2$ or 2/3 for
 radiation or matter domination respectively.  Eventually $a$ will
 asymptote to $t$ (if there were no dark energy) or to an exponential if there
 is a cosmological constant in the bubble.  We consider in detail the realistic case with arbitrary matter, radiation, curvature, and cosmological constant content in Appendix~\ref{app:conf}.  Here, we illustrate the calculation for the simpler case where the scale factor is a power law of the cosmic time.

Taking $a(t) = H_i^{-1} e^N
 (t/t_{RH})^p$ and integrating, we get
\be
\eta(t) - \eta_{RH} = {1 \over 1 - p} \left({t \over a(t)} - {t_e \over
  a(t_e)} \right)
\ee
For times much later than reheating, and many efoldings of inflation,
this is
\be
\eta(t) =  \log{\hout \over \hi} + {1 \over 1-p}{t \over a(t)}
\ee
We would like to relate this to observable quantities. The ``curvature
contribution'' to the total energy density is defined by
\be
\Omega_k(t) \equiv {1 \over H^2 a^2}  = {1 \over \dot a^2}
\ee
We have $\dot a = p a/t$
so the conformal time can be written
\be
\eta(t)  = \log{\hout \over \hi} + {p \over 1-p}\sqrt{\Omega_k(t)}
\ee
Finally, most of the time (and most of the conformal time) 
since reheating is during matter domination,
so setting $p=2/3$ we have
\be
\eta(t) \approx  \log{\hout \over \hi} + 2\sqrt{\Omega_k(t)}
\ee
valid for times well into matter domination and before dark energy or curvature domination. 
More generally we find that the expression for the conformal time today is
\be
\eta_0 \approx  \log{\hout \over \hi} + \zetao \sqrt{\Omega_k}
\ee
where $\zetao \approx 3.3$ for allowed values of the cosmological parameters (see Appendix~\ref{app:conf}).

The conformal time grows logarithmically with time for the first efold
of inflation, rapidly asymptoting to $ \log H_f/H_i$.  After
inflation it remains almost constant until late times, when---if the
curvature becomes significant---it will begin to grow again.
If our universe has a small positive vacuum energy and is nearly flat then curvature domination never
occur and the effects of collisions weaken from the onset of cosmological constant domination onward, making
the current cosmological era an opportune time to observe them
\cite{kks}.

The total number of collisions in a universe similar to ours  at the present time $t_0$ will be
\be
N \approx {8 \pi \gamma \over 3} \left( \sinh 2 \eta_0 + \eta_0
\right)  
\approx {4 \pi  \gamma \over 3} \left[ \left({H_f \over H_i}\right)^2
  e^{2\zetao \sqrt{\Omega_k(t)}} + 2 \log {\hout \over \hi} + 2\zetao
    \sqrt{\Omega_k(t)} \right] .
\ee  
To compute the total number in our backward lightcone now, we use $\Omega_k(t_0) \ll 1$ and also expand for $\hout/\hi \gg 1$ to get
\be
N \approx {4 \pi \gamma \over 3} \left({H_f \over H_i}\right)^2(1 + 2\zetao \sqrt{\Omega_k} + ...)
\ee
where the $...$ denotes terms which are much smaller than one. (See
appendix C of \cite{aj2} for a more detailed computation of this quantity.) The $\sqrt{\Omega_k}$ term is at most an order one multiplicative correction, so the bottom line is
\be
N \approx  {4 \pi \gamma \over 3} \left({H_f \over H_i}\right)^2
\ee
There is no reason to expect this number to be small.  While one expects $\gamma \ll 1$, observational constraints require $({H_f / H_i})^2 \gtrsim 10^{12}$ if $H_f \sim M_{\rm Pl}$.  In models of low-scale inflation $H_i$ can be very small; in such models the ratio $({H_f / H_i})^2$ can be $10^{80}$ or more.  Moreover, the relevant $\gamma$ is the fastest decay channel of the false vacuum around the observation bubble.  In models like the string theory landscape, there are an enormous number of decay channels available.  $\gamma<1$ is required for our analysis to be valid, but it is not clear how much smaller than one we should expect it to be.

\subsection{Distribution at Last Scattering}

One of the most promising possibilities for observing these collisions comes from the cosmic
microwave background.  In order for a collision to make an observable mark on the CMB,
the collision bubble lightcone should be of reasonable size when it intersects the part of the last 
scattering volume we can observe.  If it is smaller than a fraction of a degree it will be lost in the noise, and if it is much larger than the observable part of the reheating volume the signal will be degenerate with the dipole from the peculiar velocity of the earth in the CMB rest frame \cite{ckl2}.  However if it has an angular size of order one, it can create potentially observable disks on the CMB inside of which the average temperature is affected in an angle-dependent way \cite{ckl2}.
Therefore the collisions of most interest observationally are those with lightcones that bisect the observable part of the last scattering volume.

We describe the bubbles first in terms of their comoving distance from
$\rho=0$ at last scattering. Then we will convert this to their
angular size. For a collision which is not yet in the backward
lightcone of $\rho = 0$ at last scattering, the shortest comoving
distance to the lightcone is given by the part of the lightcone which travels radially,
\be
\rho + \etals = \eta_v
\ee
so that the comoving distance is
\be
\c (\etals) = \eta_v - \etals
\ee
For bubbles that are already in the backward lightcone of $\rho=0$ at
last scattering, $\c$ is negative but its magnitude still gives the shortest
comoving distance from $\rho =0 $ to the collision lightcone.

To summarize, we will trade in $\eta_v$, the time the future lightcone
of the collision crosses $\rho = 0$, for $\c$, the comoving distance
from $\rho = 0$ to the collision lightcone at last scattering. We allow $\c$
to be positive or negative, with positive $\c$ corresponding to
collisions that have not yet crossed $\rho = 0$ at last scattering,
and negative $\c$
corresponding to collision that have already crossed $\rho = 0$.

Let's rewrite our distribution in terms of $\c$. 
We change variables in the distribution by substituting
\be
\eta_v = \c + \etals
\ee
so the distribution is
\begin{eqnarray}
dN &=& \gamma { \cosh^2(\c + \etals - \x) \over \cosh^4 \x} d \x d \c
d^2\Omega_2 \\
0 &<& \c + \etals
\end{eqnarray}
where the lower equation arises from demanding that the collision does
not disrupt structure formation at $\rho = 0$.

We can integrate out $\x$ as before
to get
\be
dN = {2 \gamma \over 3} [1 + 2 \cosh(2(\c + \etals))] d\c
d^2\Omega_2
\label{cdist}
\ee
This gives the distribution of comoving distances to collision
lightcones at last scattering. If we were interested in the disribution
at some other time $\eta_1$, we could just replace $\etals \to \eta_1$
in this formula.

\subsection{Effects on the Last Scattering Surface}
A key measure of the observability of a given collision is the the
ratio $\rhols/\c$, where $\rhols$ is the comoving distance to the last
scattering surface and $\c$ is the comoving distance to the collision
lightcone. Recall that $\c$ is negative for collision lightcones that
have crossed the origin at last scattering and positive for collision
lightcones that have not yet crossed the origin. 

Bubbles
with $\c > \rhols$ are spacelike separated from the entire last
scattering surface and have no effect. Bubbles with $ - \rhols < \c <
\rhols$ affect only part of the last scattering surface and have the
most striking signals. Bubbles with $\c < - \rhols$ influence the
entire last scattering surface. They may lead to interesting effects,
especially at long wavelengths. However, it is clear that as the
distance to the lightcone gets much larger than the distance to the
last scattering surface,
$\c/\rhols \to - \infty$, the anisotropy induced by the collision must vanish.

 The comoving distance to the last scattering surface is related to
 the curvature,
\begin{eqnarray}
 \rho_{LS} \approx \zetao \sqrt {\Omega_k(t_0)}, 
\end{eqnarray}
where $\Omega_k(t_0)$ is the curvature contribution to $\Omega_{\rm
  tot}$ today and $\zetao \sim 3$ is a cosmological-parameter dependent factor (see Appendix~\ref{app:conf}).   Observational bounds on curvature give $\rhols \ll 1$.

As mentioned above, perhaps the most interesting bubbles are those
whose future lightcones intersect part of the reheating surface. The
number of such bubbles can be obtained by integrating the distribution
(\ref{cdist}) over the range $-\rhols < \c < \rhols$. 
The total number of such bubbles is
\be
N_{LS} = {8 \pi \gamma \over 3} [ 2 \rhols + \sinh(2 \etals + 2
\rhols) - \sinh(2 \etals - 2 \rhols)]
\ee
Using $\rhols \ll 1$ and $\etals \gg 1$
this simplifies to
\be
N_{LS} \approx {16 \pi \gamma \over 3}\rhols e^{2 \etals}
\ee
Now using $\rhols \simeq \zetao \sqrt{\Omega_k(t_0)} $ and $\etals = \log{
  \hout \over \hi}$, we get
\be
N_{LS} \approx {16 \pi \gamma \over 3}\zetao\left( \hout \over \hi \right)^2 \sqrt{\Omega_k(t_0)}
\ee
Comparing this to the total number of bubbles, we have
\be
{N_{LS} \over N} \approx 4\zetao \sqrt{\Omega_k(t_0)} \sim 12\sqrt{\Omega_k(t_0)}
\ee
valid for small $\sqrt{\Omega_k(t_0)}$.

For bubbles that intersect the last scattering surface, we can change
variables from the distance $\c$ to the angular size $\psils$ on the
CMB sky. In other words, $\psils$ is the angular radius of the part of
the last scattering surface affected by the collision. Because the
last scattering surface is small compared to the radius of curvature,
we can approximate the spatial geometry inside the last scattering
surface as flat. Also,
the collision lightcone has grown large by last scattering, so it 
is nearly a straight line as it crosses the region inside the last
scattering surface. Therefore we can use trigonometry to obtain
\be
{\c \over \rhols} = \cos \psils
\ee
Using the same approximations as above, the distribution in angles is
\be \label{psils}
dN \simeq {2\zetao \gamma \over 3}\left( \hout \over \hi \right)^2
\sqrt{\Omega_k(t_0)} d(\cos \psils) d^2 \Omega_2~.
\ee
The distribution of angular sizes is rather featureless, 
\be
dN \propto d(\cos \psils)~.
\ee

\section{Discussion}

Our primary goal in this paper was to re-consider the probability of
observing a cosmic bubble collision in a model like the string theory
landscape.  We have found that incorporating the effects of a
realistic cosmology inside the observation bubble has a dramatic
effect on the result, due to the appearance of factors of the
dimensionless ratio of the energy scale of the false vacuum to the
energy scale of inflation inside the bubble.  When this ratio is large
(as it must be absent significant fine-tuning) the number of bubble
collisions in the past lightcone of an observer in a universe like
ours is enhanced, $N \sim \gamma (H_f/H_i)^2$.
Physically this ratio arises for a simple reason: if the observation
bubble inflates with Hubble constant $H_i$, late-time observers like
us can see a region of the domain wall with surface area $\hi^{-2}$,
which is large in false vacuum Hubble units. As a result, we can see
many Hubble volumes of the false vacuum in which collision bubbles may
have nucleated.

In addition to the total number we computed the collision distribution
as a function of proper distance between nucleation points in the false vacuum, angular
location on the observer's sky, proper time when the collision first enters the observers
past lightcone, and angular size on the CMB sky (and on the wall of
the observation bubble in Appendix B).  The results are striking
in several ways.  The distribution, when projected onto the
observation bubble wall at late times, is globally conformally
invariant ({\em i.e.} invariant under the Lorentz transformations of
the nucleation point) in the limit of small angular size.  The
anisotropy in the distribution, which is present for all but a special
observer inside, turns out to be extremely small (because nearly all
of the collisions visible to such an observer have very small size on
the bubble wall, and the distribution in that limit is rotationally
invariant). In particular, as we will see in Appendix C the
anisotropic terms in the distribution are suppressed by a factor of
$(H_i/H_f)^6$.

The distribution on the last scattering surface is of particular relevance for the potential observability of these events.  Collisions with lightcones that bisect the part of the last scattering surface we can observe can create hot or cold disks on the CMB temperature map, and if these disks are of reasonable size and contrast they could be detectable \cite{ckl2}.  We find that the number of such disks is $N_{LS} \sim \sqrt{\Omega_k} N\sim \gamma (H_f/H_i)^2 \sqrt{\Omega_k}$.  Hence $N_{LS}$ scales as $\sqrt{\Omega_k} = 1/{\dot a} \sim e^{-(N-N^*)}$, where $N^*$ is the number of efolds of inflation needed to give the universe a radius curvature of order .1 in Hubble units (as required by current observational bounds).  

As expected, long inflation inflates away the effects of bubble collisions by inflating away their number density---but since the experimental constraint on curvature is not very strong, the total number can still be quite large with no fine-tuning.  However even if $N_{LS} >1$, one must also consider the effects of inflation on the magnitude of the signal from each collision.  The analysis of \cite{ckl2} showed that a single collision with $N-N^* \sim 5$ can have an easily observable effect without fine-tuning the parameters of the collision bubble itself or our boost towards it (which is determined by $\x$ in our notation).  If $N_{LS}$ is large or if some of the collisions are with bubble types that have particularly dramatic effects on the early universe, $N-N^*$ could be considerably larger than this and still produce observable effects on the CMB temperature map.  

The results of this analysis are interesting for a number of reasons.  For one, an observation of the effects of a collision of this type would be, if not a direct confirmation of the string theory landscape, at least very dramatic evidence in its favor.  The observational signatures of these collisions has been studied in detail  in the context of a collision between two bubbles in a number of recent papers \cite{ckl1, aj2, ckl2, aj3}, and if they are absent in the data this could constitute an interesting constraint on string models.

Our findings open a large array of directions for future research.  One important issue is to make more explicit the connection between the thin-wall Coleman de-Luccia bubbles considered here and the instanton transitions in the string theory landscape itself.  Another is to further study the effects of the collisions on inflation and the CMB sky, and to investigate additional observational opportunities like the effects on large scale structure or CMB polarization.  A particularly interesting possibility along those lines is cosmic 21 cm data, which has tremendous potential as a probe of high-scale physics in the early universe \cite{Loeb:2003ya, Kleban:2007jd,Mao:2008ug}.  It would be very useful to run a numerical simulation to verify the validity our assumptions about the domain wall dynamics and their effects of the reheating surface.  There is an intriguing connection to conformal field theory, which will be explored in \cite{fk}. Finally, there remains the larger issue of the probability measure in the full string theory landscape.

\paragraph{Appendices:}

In Appendix A, we address the issue of bubbles nucleating inside bubbles with a rate different than that of the parent false vacuum.  We find that even in the extreme case that all bubbles are absolutely stable (so that no nucleations occur inside them), our distribution is unchanged up to factors of order $\gamma \ln (H_f/H_i)$, which is generically small.

In Appendix B we compute the distribution of collisions on the asymptotic boundary of the observation bubble using a slightly different technique than that in the bulk of the paper.  The results are consistent up to the expected accuracy,  and we make the connection to the results in \cite{ggv} more explicit. 

In Appendix C we quantify in detail the level of anisotropy in the collision distribution.

In Appendix D we derive an expression for the conformal time including the effects of early-time radiation domination and late-time dark energy domination.  

\section*{Acknowledgements}

We would like to thank Puneet Batra, Andrei Gruzinov, Lam Hui, Andrei Linde, Massimo Porrati, Steve Shenker, Lenny Susskind, and Alex Vilenkin for useful discussions.  BF would like to thank Raphael Bousso and I-Sheng Yang for extensive discussions and collaboration on topics very closely related to the work presented here.
The work of MK is supported by NSF CAREER grant PHY-0645435.   The work of KS is supported in part by a Natural Sciences and Engineering Research Council of Canada Discovery Grant.
We thank the Aspen Center for Physics, where this work was initiated, for its hospitality.

\appendix

\section{Proper treatment of overlapping bubbles}
The analyses in the paper did not take into account the effects of the interactions of bubble
nucleations with each other.
After a collision bubble has nucleated, its future lightcone is no
longer available for the same type of nucleation event. The vacuum
inside the collision bubble may be completely stable; in any case it will certainly
have a different set of decay rates than that of the parent false vacuum. As a simple test to see how significant this effect can be for our analysis we
will assume the bubbles are absolutely stable, and therefore eliminate all bubbles from our distribution that would have nucleated inside the future lightcone of
other bubbles. 

 Given
two points $A$ and $B$ in de Sitter space, we need a formula for when
$A$ and $B$ are spacelike separated. In terms
of the embedding coordinates, the condition is
\be
\eta_{\mu \nu} X^\mu_A X^\nu_B = X_A X_B + R_A R_B \cos (\delta
\theta) - T_A T_B < 1
\ee
where $\delta \theta$ is the angular separation between the centers of
the two bubbles. Changing to our coordinates, the condition is
\be
\cosh (\t_A - \t_B) < \cosh(\x_A - \x_B) + \cosh \t_A \cosh \t_B (1 -
\cos (\delta \theta))
\label{flc}
\ee

To take this restriction into account, one could distribute bubbles
according to (\ref{dist}) and then eliminate any bubble that is
in the future lightcone of another bubble.
However, we will show that the expected number of bubbles which must be
eliminated is generically much less than one, so that this correction can be
neglected. 

To show this, we consider the $S_2$ defined by the
intersection of the
observer's backward lightcone with the domain wall of the observation
bubble. In coordinates, this $S_2$ is defined by 
$\eta = - \infty$, $ \rho = \infty$, $\eta + \rho = \eta_0$. Since
this $S_2$ is on the lightcone of the observation bubble, it can also
be defined in the coordinates that cover the exterior of the bubble
as the surface $\x = - \infty$, $\t = \infty$, $\x + \t = \eta_0$.

A given collision bubble affects a solid angle $\Omega$ on this
$S_2$. We can use the formula (\ref{flc}) to find that the solid angle
inside the future lightcone of a bubble with coordinates $(\x, \eta_v)$ is
\be
\Omega = 2 \pi { e^\x(e^{-\eta_v} - e^{-\eta_0}) \over \cosh(\eta_v - \x)}
\ee
while the angular location of the bubble is randomly distributed.

A sufficient (but not necessary) condition for two nucleation events to be spacelike
separated is that the solid angle on the observation bubble's wall affected by one is not completely
contained in the solid angle affected by the other. Therefore, if the
expected total solid angle affected by all of the bubbles is much less
than $4 \pi$, then it is unlikely that any bubble is in the future
lightcone of another. The expected total solid angle is
\be \label{area}
\langle {\Omega \over 4 \pi} \rangle = \gamma \int_0^{\eta_0} d \eta_v
\int_{-\infty}^\infty d\x \int d^2\Omega {\cosh^2(\eta_v - \x) \over
  \cosh^4 \x}  { e^\x(e^{-\eta_v} - e^{-\eta_0}) \over 2 \cosh(\eta_v -
  \x)} = \gamma \left( {\eta_0 \over 3} + {\cal O}(e^{-\eta_0}) \right).
\ee
As we have seen, for a realistic cosmology $\eta_0 \sim \ln {H_f \over H_i}$.  Since $\gamma$ is exponentially small,  $\gamma\eta_0$ will generally be much less than 1.

\section{Collision distribution on the asymptotic sphere}

In this appendix we will compute the bubble collision distribution  as a function of the angular radius on the asymptotic boundary sphere---which is the boundary of any open spatial slice inside the observation bubble, or equivalently the observation bubble's lightcone after infinite time---taken up by collisions.  This in general is not the same as the observed angular size of a bubble as seen by a finite time observer. In fact, not all the bubbles that chop off  a portion of the boundary are going to be visible to any given observer even after infinite time unless $\Lambda_{\rm in } =0 $.  Hence,  the distribution we are going to compute here is not a directly observable quantity. Nevertheless we find it a useful concept, as it characterizes the statistics of collision events unambiguously, and independently of the observer and of the Hubble rate inside our bubble. Indeed, precisely because we are not keeping track of the propagation of the collision signal inside our bubble, the distribution we get cannot depend on the geometry of our bubble, but only on the geometry of its wall---which we approximate as the future light-cone of the nucleation event---and of the outside. As a result, this distribution will be applicable to any cosmology inside.  (For the distribution actually seen by a realistic observer, see the analysis of Section 4.)

\begin{figure}[b!]
\begin{center}
\includegraphics[width=6cm]{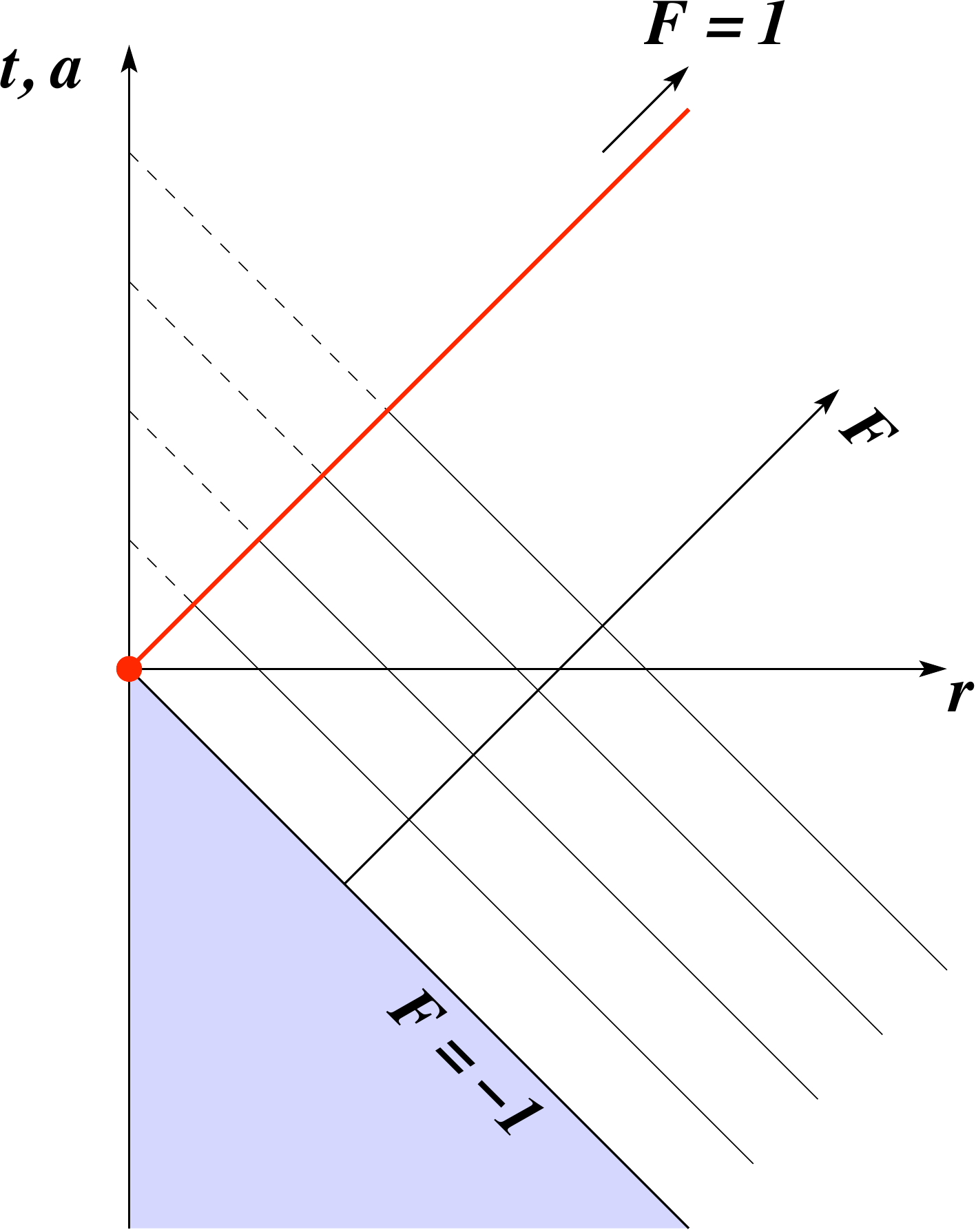}
\end{center}
\caption{\label{aF} \small The $(a, F)$ coordinate system described in the text. Our bubble nucleates at the origin $(t=0, r=0)$.}
\end{figure}

For simplicity we set $H_{f} =1$. The dependence on $H_{f}$ can be recovered at any stage of our computation just by dimensional analysis. In flat FRW coordinates the metric outside our bubble is
\be \label{metric_rt}
ds^2 = dt^2 - e^{2t} \big( dr^2 + r^2 d\Omega^2 \big) \; .
\ee
It is actually more convenient to use the $(a,F)$ coordinatization defined by \cite{ggv},
\be
a = e^t \; , \qquad F = r - e^{-t} \; ,
\ee
for which the metric reads
\be \label{metric_aF}
ds^2 = 2 \, da dF - a^2 dF^2 - (1+aF)^2 d\Omega^2 \; .
\ee
$a$ is just the scale factor, while the constant-$F$ hypersurfaces are past directed light-cones, with tip at $r=0$ (see fig.~\ref{aF}). This chart can in fact cover all of deSitter space, if we let $a$ and $F$ take any real values, with the constraint $aF \ge -1$ ($r \ge 0$). However we will be interested in a smaller range of $a$ and $F$. First, the surface where initial conditions are given  (which we take following \cite{ggv}) corresponds to $a=0$.  Then, our bubble's nucleation point, $t=0, r=0$, corresponds to $a = 1, F = -1$. Therefore, its past light-cone is the surface $F = -1$, with $0< a < 1$.
On the other hand, our bubble's wall---the future-directed light-cone with tip at $(t=0, r=0)$---is the surface $a = 2/(1-F)$, with $-1 < F < 1$. The region outside both these light-cones, where other nucleation events can take place {\em and} eventually lead to collisions with our bubble is
\be \label{range_aF}
-1 < F < 1 \; , \qquad 0 < a < \frac{2}{1-F} \; .
\ee
Notice in particular that $F < 1$ is necessary for the bubble nucleated at $(a,F)$ to eventually collide with ours.

Consider then a bubble nucleated at $(a, F)$. Its angular size $\psi$ on the boundary sphere is given by \cite{ggv}
\be \label{angular_size}
\mu \equiv \cos \psi = \frac{1}{2} \frac{a+2F +aF^2}{1+aF} \; .
\ee
Since in the end we are interested in the distribution as a function of $\mu$, we can change variables and trade the nucleation $a$ for $\mu$,
\be \label{a}
a = \frac{\mu - F}{1+F^2 -2F \mu} \; .
\ee 
This way each collision event will be labeled by its angular size $\mu$ and by $F$, which is `conserved' between nucleation and collision---because the bubble wall follows a null trajectory with $F = {\rm const}$. The allowed ranges for these variables are
\be \label{range_muF}
-1 < \mu < 1 \; , \qquad -1 < F < \mu \; .
\ee
Indeed eq.~(\ref{angular_size}) is an increasing function of $a$ for fixed $F$. This means that the maximum and minimum of $\mu$ are attained at the extrema of $a$, for a given $F$. This yields the ranges above.
The infinitesimal four-volume element corresponding to the infinitesimal intervals $d \mu$, $dF$ is
\be
dV_4= 4 \pi (1+aF)^2 \, da \, dF = \frac{8\pi \, (1-F^2)^3}{(1+F^2 -2F \mu)^4} \, d\mu \, dF \; ,
\ee
where we used eq.~(\ref{a}) for the change of variable $a \to \mu$, and the $4\pi$ comes from the solid angle integral. Then, the four-volume where a bubble of size between $\mu$ and $\mu+d\mu$ can nucleate is just the integral of this along $F$, from $-1$ to $\mu$ (see eq.~(\ref{range_muF})),
\be
\frac{dV_4}{d \mu} = \int_{-1} ^ \mu \! dF \, \frac{dV_4}{d \mu dF} = \frac{4 \pi}{3} \frac{(2-\mu)}{(1-\mu)^2} \; .
\ee
A bubble of size $\mu$ takes out a solid angle $2\pi(1-\mu)$ on the boundary sphere. The solid angle $\Omega_f(\mu)$ that is free of bubbles with angular sizes from $\mu = -1$ to $\mu$ obeys the equation
\be
\frac{d \Omega_f(\mu) }{d \mu} = -2\pi(1-\mu) \cdot \gamma \cdot \frac{\Omega_f(\mu)}{4\pi}  \frac{dV_4 }{d \mu} \; ,
\ee
where $\gamma$ is the nucleation rate, and we included the factor $\Omega_f / 4\pi$ to  account for the fact that at any given $\mu$ only such a fraction of the solid angle is still available for new nucleations and collisions (in other words, we are counting regions in which disks overlap only once, no matter how many such overlaps there may be) . The solution with boundary condition $\Omega_f(-1)= 4\pi$ is 
\be \label{Omega_f}
\Omega_f (\mu) = 4 \pi \left (\frac{1-\mu}{2} \right)^{\frac{2}{3}\pi \gamma} e^{- \frac{2}{3}\pi \gamma(1+\mu)}
\ee
or, in differential form,
\be \label{dOmegaf}
\frac{d \Omega_f(\mu) }{d \mu} = - \frac{8\pi^2 \gamma}{3} \,  \frac{2-\mu}{1-\mu} \left (\frac{1-\mu}{2} \right)^{\frac{2}{3}\pi \gamma} e^{- \frac{2}{3}\pi \gamma(1+\mu)}
\ee
As a check notice  that if we consider all bubble sizes, from $\mu = -1$ to $\mu =1$, all the solid angle but a set of measure zero is eaten up by collisions,
\be
\Omega_f (1) = 0 \; ,
\ee
as expected.
Another quantity of interest is the expected number of bubbles with angular size between $\mu$ and $\mu+d\mu$,
\be \label{dn}
\frac{d N(\mu)}{d \mu} = -\frac{\gamma}{2\pi(1-\mu)} \frac{d \Omega_f(\mu) }{d \mu} \; .
\ee
Integrating this we get total number of bubbles down to angular size $\mu$:
\bea
N(\mu) & = & \frac{4 \pi \gamma}{3} \left( -\frac{3}{4 e^2  \pi \gamma} \right)^{\frac{2\pi\gamma}{3}}
\bigg[  \gamma\big( \sfrac{2\pi\gamma}{3}, - \sfrac{2\pi\gamma}{3}(1-\mu),  -\sfrac{4\pi\gamma}{3}\big) \nn \\
&& \qquad \qquad - \frac{2\pi\gamma}{3} \gamma\big( -1+\sfrac{2\pi\gamma}{3}, - \sfrac{2\pi\gamma}{3}(1-\mu),  -\sfrac{4\pi\gamma}{3}\big) \bigg] \; ,
\eea
where $\gamma(a,z_1,z_2)$ is the generalized incomplete gamma function, $\gamma(a,z_1,z_2) \equiv \int_{z_1} ^{z_2} t^{a-1} e^{-t} dt$.
Perhaps more interesting is the asymptotic behavior of $n(\mu)$ at small angular sizes $\psi \ll1$. Expanding $\mu = \cos \psi$ we get
\be \label{n}
N(\psi) = \frac{8\pi \gamma}{(2-\frac{2\pi\gamma}{3}) (2e)^{\frac{4\pi\gamma}{3}}} \cdot \frac{1}{{\psi}^{2-\frac{4\pi\gamma}{3}}} + {\cal O}\big( 1 \big) \; ,
\ee
which gives us the fractal dimension of our ensemble, $d= 2-\frac{4\pi\gamma}{3}$, in agreement with GGV's result \cite{ggv}.

Although we did not keep track of the bubble positions on the boundary sphere,
the distribution we have computed is obviously isotropic, because we have computed it in a frame where the initial-condition surface ($a=0$) is invariant under rotations. This is the distribution that could in principle be seen by the central observer after infinite time, if she lives in a bubble of Minkowski vacuum and can see all the way to her bubble's domain wall. To compute the boundary distribution relevant for a different observer, we have to ``boost'' eq.~(\ref{Omega_f}). 

\subsection{Trasformation properties under boosts}

We are thus led to consider how a deSitter boost acts on the boundary sphere. Recall that deSitter boosts about our bubble's nucleation point (henceforth ``the origin'') act as spatial translations on any given hyperbolic FRW slice inside the bubble. That is why different comoving FRW observers are mapped into each other by boosts about the origin. However since we are interested just in the boundary sphere, we can use the fact that the isometries of three-dimensional hyperbolic space acts as the two-dimensional conformal group on the boundary. In particular, a translation (= deSitter boost) along $z$ acts as a dilation about the south pole and a contraction about the north pole. More precisely,
consider the metric on the sphere
\be
d \Omega_2^2 =  d \theta^2 + \sin ^2 \theta \, d \varphi^2 \; ;
\ee
it is a conformally flat metric, as made evident by the change of variable
\be \label{stereo}
\rho = \frac{1+\cos \theta}{\sin \theta} \qquad \leftrightarrow \qquad \sin \theta = \frac{2 \rho}{1+\rho^2} \; ,
\ee
which is the standard stereographic projection, mapping the sphere to the plane. The North Pole (NP) $\theta = 0$ gets projected to infinity, whereas the South Pole (SP) $\theta = \pi$ gets mapped to the origin $\rho =0 $.
Upon the stereographic projection the metric becomes
\be
d \Omega_2^2 = \frac{4}{(1+\rho^2)^2} \big(  d \rho^2 + \rho^2 \, d \varphi^2\big) \; .
\ee
Then, dilations about the origin, $\rho \to \lambda \rho$, leave the metric explicitly conformally flat. These are the trasformations that correspond to translations along $z$ in the three-dimensional hyperbolic interior of the sphere.
In terms of $\theta$ and $\varphi$ they act as
\be \label{conf_transf}
\sin \theta \to \sin \theta' = \sin \theta \frac{2 \lambda}{(\lambda^2+1) + 
(\lambda^2-1) \cos \theta} \; , \qquad \varphi \to \varphi' =\varphi \; .
\ee
The action of this transformation is schematically depicted in fig.~\ref{boosts}.

\begin{figure}[t!]
\begin{center}
\includegraphics[width=13cm]{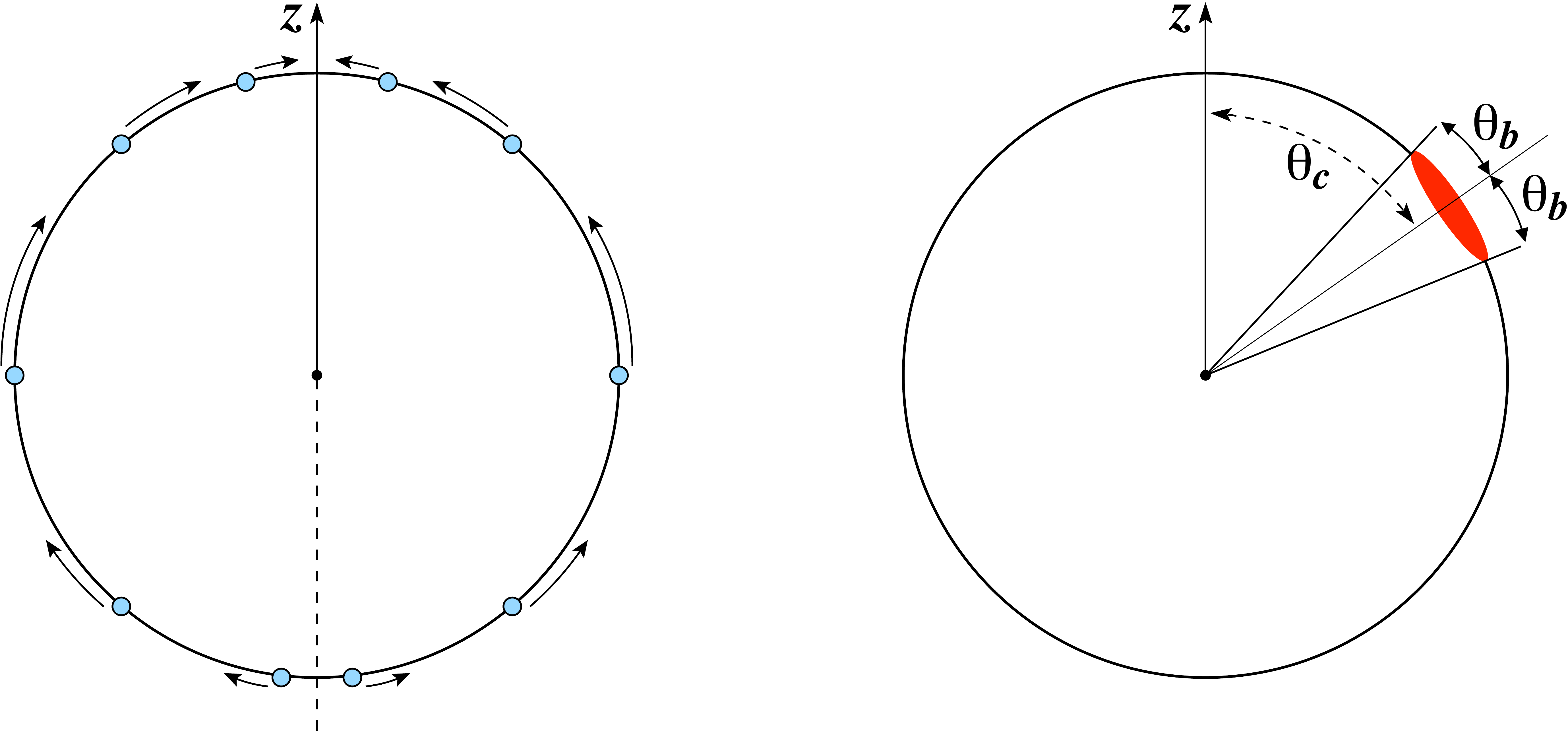}
\end{center}
\caption{\label{boosts}  \small 
{\em Left:} A boost along $z$ acts on the boundary sphere as a dilation about the south pole and a contraction about the north pole.
{\em Right:} A circle on the boundary sphere is parameterized by its center's coordinates $(\theta, \varphi)$ and its angular radius $\psi$.
}
\end{figure}

Bubble collisions chop off circles on the boundary sphere. Interestingly enough, circles get mapped into circles by our conformal transformation eq.~(\ref{conf_transf}). To see this, recall that the stereographic projection (\ref{stereo}) maps circles on the sphere to circles on the plane. Combining this with a  dilation on the plane $\rho \to \lambda \rho$, which clearly maps circles into circles, and with an inverse stereographic projection, we get that eq.~(\ref{conf_transf}) maps circles to circles.
So, given a circle we just have to understand what happens to its center and its radius upon the transformation. Notice that the new center's $\theta$ will not be simply the transformed $\theta$ of the old center according to eq.~(\ref{conf_transf}). Rather, given a bubble with center at $\theta$ and angular radius $\psi$ (see fig.~\ref{boosts}), to find the new center's position we have to transform the two extrema of the bubble along theta, $(\theta \pm \psi)$, and take their half-sum,
\be \label{boosted_position}
\theta ' \equiv \frac{(\theta + \psi)' + (\theta- \psi)'}{2} \; .
\ee
Likewise, the new angular radius is given by
\be \label{boosted_size}
\psi ' \equiv \frac{(\theta + \psi)' - (\theta- \psi)'}{2}
\ee

To simplify the algebra, we can consider two limiting cases: that of an infinitesimal transformation, $\lambda = 1+ \epsilon$, and that of an infinite one, $\lambda \gg 1$. In the former case we have an infinitesimal, $\theta$-dependent shift on $\theta$,
\be
\delta \theta \equiv \theta' - \theta \simeq - \epsilon \sin \theta \; .
\ee
For a bubble centered at $\theta$ with angular radius $\psi$ we get 
\be
\theta \simeq  \theta - \epsilon \sin \theta \cos \psi \; , \qquad
\psi '  \simeq \psi - \epsilon \sin \psi \cos \theta \; . 
\ee
If we restrict to very small bubbles, $\psi \ll 1$, these transformation laws further simplify to 
\be \label{simple_transf}
\theta \simeq  \theta - \epsilon \sin \theta  \; , \qquad
\psi '  \simeq \psi (1 - \epsilon  \cos \theta) \; . 
\ee
The number of bubbles per solid angle down to angular size $\psi$ is given by eq.~(\ref{n})
\be \label{dnint}
dN \propto \frac{\gamma}{{\psi}^{2-\frac{4\pi\gamma}{3}}}  \, d (\cos \theta) \; ,
\ee
where the $d (\cos \theta)$ comes from the isotropy of our distribution. The number $dn$ is invariant under our boost, whereas $\theta$ and $\psi$ transform as in eq.~(\ref{simple_transf}). Therefore the boosted distribution reads
\be
dN \propto \frac{\gamma}{{\psi'}^{2-\frac{4\pi\gamma}{3}}}  \, \big(1+ \epsilon \, \sfrac{4\pi\gamma}{3} \cos \theta ' \big) \, d (\cos \theta ') \; .
\ee
The boost introduces a dipolar anisotropy in the bubble-collision distribution.

More interesting for us is the opposite limit, that of infinite boost $\lambda \to \infty$. To figure out its action, it is best to work directly on the stereographic plane. Given a circle on the sphere with center at $\theta$ and angular size $\psi$, this corresponds to a circle on the plane with center at $\rho_c$ and radius $\rho_b$, 
\be
\rho_c = \sfrac12(\rho_+ + \rho_-) \; , \qquad \rho_b = \sfrac12(\rho_+ - \rho_-) \; ,
\ee
where $\rho_\pm$ are the projections of $\theta_{\pm} \equiv \theta \pm \psi$, according to eq.~(\ref{stereo}).
After some some straightforward algebra we get
\be \label{direct}
\rho_c = \frac{\sqrt{1-x^2}}{\mu - x} \; , \qquad \rho_b = \frac{\sqrt{1-\mu^2}}{\mu - x}  \; ,
\ee
where we have defined
\be
x \equiv \cos \theta \; , \qquad \mu = \cos \psi \; .
\ee
The inverse mapping is
\be \label{inverse}
x= \frac{\rho_c^2 - \rho_b^2 -1}{\sqrt{(\rho_c^2 + \rho_b^2 +1)^2- 4 \rho_c^2 \rho_b^2}}
\;, \qquad
\mu = \frac{\rho_c^2 - \rho_b^2 +1}{\sqrt{(\rho_c^2 + \rho_b^2 +1)^2- 4 \rho_c^2 \rho_b^2}} \; .
\ee

Now, the number of collisions per unit $x$ and $\mu$ is (see eqs.~(\ref{dn}, \ref{dOmegaf}), and recall that before boosting the distribution is isotropic)
\be
dN \sim \frac{1}{1-\mu} \, d \Omega_f \sim \gamma \frac{2-\mu}{(1-\mu)^2}\left (\frac{1-\mu}{2} \right)^{\frac{2}{3}\pi \gamma} e^{- \frac{2}{3}\pi \gamma(1+\mu)} \, dx \, d\mu \; ,
\ee
which in terms of $\rho_c$ and $\rho_b$ reads
\be
dN \sim \gamma \left[ (2- \mu)(1+\mu)^2 \left (\frac{1-\mu}{2} \right)^{\frac{2}{3}\pi \gamma} e^{- \frac{2}{3}\pi \gamma(1+\mu)}\right] \times \frac{\rho_c}{\rho_b^3} \, d \rho_b \, d \rho_c  \; .
\ee
The piece outside the brackets is manifestly invariant under dilations $\rho \to \lambda \rho$, which are precisely the transformations we are interested in. We just have to figure out what happens to the terms in brackets.
Performing a very large boost corresponds to using a new radial coordinate
\be
\rho' = \lambda \rho \; , \qquad \lambda \gg 1 \; .
\ee
Finite values of $\rho'$ correspond to tiny values of $\rho$, which on the sphere correspond to very small displacements from the SP, $x \simeq -1$ and $\mu \simeq 1$. More precisely, from eq.~(\ref{inverse})
 we have
 \be
 \mu \simeq 1 - \frac{2}{\lambda^2} \rho'_b {}^2 \; .
 \ee
Plugging this into our distribution and dropping the primes we get
\be
dN \sim \frac{\gamma}{\lambda^{\frac{4}{3}\pi \gamma}} \frac{\rho_c}{\rho_b {}^{3-\frac{4}{3}\pi \gamma}} \, d \rho_b \, d \rho_c
\ee
which projected back onto the sphere via eq.~(\ref{direct}) reads
\be \label{Adist}
dN \sim  \frac{\gamma}{ \lambda^{ \frac{4}{3}\pi \gamma} } 
\frac{d \psi} { (\sin \psi)^{ 3-\frac{4}{3}\pi \gamma } } 
\frac{d \cos \theta}{ (\cos \psi - \cos \theta)^{ \frac{4}{3}\pi \gamma }  } \; .
\ee

To see the relation to the distribution derived in the bulk of the paper (Eq.~(\ref{dist})), 
 we need to integrate the latter over $\x$.  If we do not impose any restriction on the observer from domain walls, the limits of 
 integration are from $(-\infty, \infty)$, giving (c.f. Eq.~(\ref{dist}))
\be
dN = {8 \pi \over 3} \gamma \cosh^2 \tau d \tau d \cos \theta \; .
\ee
where we have used $\etav - \x = \t$.
The relation between the coordinate $\tau$ and the asymptotic angular radius $\psi$ is simply $\tanh \tau = \cos \psi$; with a little algebra this becomes
\be \label{Bdist}
dN = {8 \pi \gamma \over 3} {d \psi \over \sin^3 \psi} d\cos \theta \; .
\ee

So far we have placed no restrictions on the direction of the boost.  However if we require that the boost avoid bubble domain walls, the distribution Eq.~(\ref{Adist}) will be modified by a factor of $\Theta(\theta - \psi)$, which enforces that no disks cover the point $\theta=0$ the boost was taken towards.  This condition coincides precisely to the one imposed by the initial condition surface at infinite boost discussed in the text below Eq.~(\ref{range}), which is $\cos \psi + \tanh \tau >0$, or in these coordinates $ \theta > \psi$.

Expanding Eq.~(\ref{Adist}) in powers of $\gamma$ gives
\be
dN \sim  \gamma {d \psi \over \sin^3 \psi} d\cos \theta \left(1 + {\cal O} \left( \gamma \ln \delta \right) \right),
\ee
in agreement with Eq.~(\ref{Bdist}) up to factors of order $\gamma \ln \delta$.  Here $\delta$ is a UV cutoff applied both to the minimum size $\psi_{\rm min} \sim \delta$ of the disks and to the boost parameter $\lambda_{\rm max} \sim 1/\delta$.  

The cutoff on boosts is necessary because Eq.~(\ref{Adist}) is an average derived under the assumption that no bubbles nucleate inside other bubbles.  Boosting the resulting distribution
infinitely leads to a singular result, since the average was taken before the boost and all but a set of measure zero of such boosts are in directions
already ``eaten" by a bubble domain wall. 

 A cutoff on $\psi$ corresponds to a cutoff on the observer's conformal time $\eta_0$ through $\x + \tau = \eta_0$.  After integrating over $\x$ this translates into $- \ln \psi_{\rm min} \sim \eta_0$, or $\psi_{\rm min} \sim H_i/H_f$.  As a quick check, according to Eq.~(\ref{Adist}) the fraction of the asymptotic sky covered by disks is then $\gamma \int \psi^2 d\psi / \sin^3 \psi \sim - \gamma \ln \psi_{\rm min} \sim \gamma \eta_0$, in agreement with Eq.~(\ref{area}).  Hence, the expansion in $\gamma$  shows that the distributions  (\ref{Adist}) and (\ref{Bdist}) agree as expected up to corrections of order $\gamma \eta_0 \approx \gamma \ln (H_f/H_i)$.

\section{Legendre moments of the collision distribution}

In this appendix we derive the Legendre moments of the bubble collision direction distribution, and compare this with the distribution found in Ref.~\cite{ggv}.   We focus on the effects of inflation with $H_i \neq H_f$ inside the bubble, but these results can be straightforwardly generalized to include arbitrary cosmological evolution inside the bubble.  Using the restriction of Eq.~({\ref{eqn:4volrestriction}) we can show that for an internal de Sitter space with $H_i \neq H_f$ the four-volume available for the nucleation of bubbles in the $\xi \rightarrow \infty$, infinite boost, limit is 

\begin{align}
\frac{d V_4}{d\time d\Omega} = &\frac{1}{3\H^3 \left[\H^2\cosh^2\left(\frac{\Htime}{2}\right)-\sinh^2\left(\frac{\Htime}{2}\right)\right]} \times \nonumber \\ &\left[
 \left( \frac{\H^2\sinh (\Htime)}{\H^2 (1-\cos \theta) \cosh^2\left(\frac{\Htime}{2}\right)+(1+\cos\theta)\sinh^2\left(\frac{\Htime}{2}\right)}\right)^3 
 -\tanh^3 \left(\frac{\Htime}{2}\right)\right]
 \end{align}
where we have adopted the notation ${\cal H}_i = {H_i}/{H_f}$ for the dimensionless ratio of Hubble rates.  In the limit $\H \rightarrow 1$ this reduces to Eq.~(45) of Ref.~\cite{ggv}. 

We define
\begin{equation}
\mu=\cos \theta \, ,
\end{equation}
and write
\begin{equation}
\frac{d V_4}{d\time d\Omega} = \frac{\alpha}{(\beta-\mu)^3}-\lambda
\label{eqn:generalvol}
\end{equation}
where
\begin{equation}
\alpha = \frac{\H^6 \sinh^3\left({\Htime}\right)}{3\H^3 \left[\H^2\cosh^2\left(\frac{\Htime}{2}\right)-\sinh^2\left(\frac{\Htime}{2}\right)\right]} 
\end{equation}

\begin{equation}
\beta = \frac{\H^2\cosh^2\left(\frac{\Htime}{2}\right)+\sinh^2\left(\frac{\Htime}{2}\right)}{\H^2\cosh^2\left(\frac{\Htime}{2}\right)-\sinh^2\left(\frac{\Htime}{2}\right)} 
\end{equation}
and
\begin{equation}
\lambda = \frac{\tanh^3\left(\frac{\Htime}{2}\right)}{3\H^3 \left[\H^2\cosh^2\left(\frac{\Htime}{2}\right)-\sinh^2\left(\frac{\Htime}{2}\right)\right]^4} 
\end{equation}



We expand Eq.~{\ref{eqn:generalvol} in orthonormal Legendre polynomials that satisfy
\begin{equation}
\int_{-1}^{1} d\mu {P}_n(\mu){P}_m(\mu) = \delta_{mn}
\end{equation}
as
\begin{equation}
\frac{d V_4}{d\time d\Omega} = \sum_{n=0}^{\infty} v_n P_n(\mu) \, .
\label{eqn:vol_expanded}
\end{equation}

The generating function for the ${P}_n(\mu)$ is
\begin{equation}
Z(\mu,u)= \sum_{n=0}^{\infty} \sqrt{\frac{2}{2n+1}} P_n(\mu) u^n = \frac{1}{\sqrt{1-2 u \mu + u^2}}
\end{equation}
so that
\begin{equation}
P_n(\mu) = \frac{\sqrt{n+\frac{1}{2}}}{n!} \left. \frac{\partial^n Z}{\partial u^n} \right|_{u=0}
\end{equation}
From $Z(\mu,t)$ we can define a generating function for the Legendre moments $v_n$ as
\begin{align}
{\cal V}& = \int_{-1}^{1} d\mu\, Z(\mu,u) \frac{d V_4}{d\time d\Omega} = \alpha\int_{-1}^{1} d\mu \frac{Z(\mu,u)}{(\beta-\mu)^3} - 2\lambda \\
 &=\alpha\left[Z_{\beta}^2 \left(\frac{2\beta-u(\beta^2+1)}{(\beta^2-1)^2}\right)+Z_{\beta}^4 \frac{3 u(\beta u-1)}{\beta^2-1}+3Z_{\beta}^5 u^2\arcoth \left[ Z_{\beta}(\beta-u) \right] \right]-2\lambda \, ,
\end{align}
where $Z_{\beta}=Z(\beta,u)$, such that
\begin{align}
v_n= \frac{\sqrt{n+\frac{1}{2}}}{n!}\left. \frac{\partial^n {\cal V}}{\partial u^n} \right|_{u=0} \, .
\end{align}

The first few moments are
\begin{align}
v_0=\frac{\sqrt{2} \alpha \beta}{(\beta^2-1)^2}-\lambda
\end{align}
\begin{align}
v_1=\frac{\sqrt{6} \alpha}{(\beta^2-1)^2}
\end{align}
\begin{align}
v_2=\frac{\sqrt{5}\alpha (5\beta-3\beta^3+3(\beta^2-1)^2 \arcoth\,\, \beta)}{\sqrt{2}(\beta^2-1)^2}
\end{align}
in the limit $\Htime \gg 1$ we have 
\begin{align}
v_0=\frac{2\sqrt{2}}{3}\left[\frac{2}{\H^3}+\frac{1}{\H}\right]e^{-\Htime} + O(e^{-2\Htime})
\end{align}
\begin{align}
v_1=\frac{2\sqrt{6}}{3}\left[\frac{1}{\H}\right]e^{-\Htime} + O(e^{-2\Htime})
\end{align}
\begin{align}
v_2 = \frac{2\sqrt{10}}{3}\left[\frac{-1+8\H^2 +12\H^4\ln \H^2 -8\H^6 + \H^8}{\H(\H^2-1)^4}\right]e^{-\Htime} + O(e^{-2\Htime})
\end{align}
from which we can see the dipole and higher multipoles are suppressed relative to the monopole by a factor of $\H^2$ in the limit that $\H$ is a small parameter, and the distribution is nearly isotropic.

To leading order in the $\H$ expansion we can write for an arbitrary multiple moment
\begin{align}
v_n=(-1)^{n+1}\frac{2\sqrt{4n+2}}{3\H}\left[1-\H^2(n+2)(n-1) + O(\H^4)\right] e^{-\Htime}+O(e^{-2\Htime})
\end{align}
from which we see that all multipole moments are present, albeit suppressed by a factor of $\H^2$, at late times when $\H \neq 1$ rather than the late-time monopole-plus-dipole distribution found in Ref.~\cite{ggv} for the $\H = 1$ case.

The late time distribution takes the exact form

\begin{align}
\frac{d V_4}{d\time d\Omega} = &\frac{4}{3\H^3 (1-\H^2)} \left[1-\frac{8\H^6}{\left[1+\cos\theta+\H^2(1-\cos\theta)\right]^3}\right]e^{-\Htime}+O(e^{-2\Htime})
 \end{align}
which in the limit $\H \rightarrow 1$ is simply
\begin{align}
\frac{d V_4}{d\time d\Omega} = 2(1+\cos\theta)e^{-\time}+O(e^{-2\time})
 \end{align}
in agreement with Ref.~\cite{ggv}.

To leading order in $\H$ the late-time distribution is
\begin{align} \label{kdist}
\frac{d V_4}{d\time d\Omega} = &\frac{4}{3\H^3} \left[1+\H^2+\H^4+ \left(1-\frac{8}{\left[1+\cos\theta+\H^2(1-\cos\theta)\right]^3}\right)\H^6 + O(\H^8)\right]e^{-\Htime}+O(e^{-2\Htime})
 \end{align}
and so the anisotropic part of the distribution is suppressed by $H_i^6$ except in the region $\pi - \theta < \epsilon$, where $\epsilon \sim H_i/H_f$.  Only a fraction of order $H_i^2/H_f^2$ of the collisions are centered within this region.

\section{Computation of the conformal time with a $\Omega_{\Lambda}\neq \Omega_{k} \neq \Omega_{r} \neq 0$ }
\label{app:conf}

In this appendix we derive an expression for the conformal time taking into account early time radiation domination and late time dark-energy domination (modeled here as a cosmological constant) of our Universe.  In this paper we use a convention where the scale factor has units of length and the curvature $k=-1$ is a dimensionless quantity.  We will not set $k=-1$ in this derivation to allow easy comparison to the alternative convention often used where the scale factor is dimensionless and $k$ has units of inverse length squared.  For a scale factor $a(t)$, where $t$ is cosmological time, the dimensionless conformal time today since some early time $t_i$ is

\begin{align}
\Delta\eta = \eta_0-\eta_e=\sqrt{-k}\int_{t_i}^{t_0}\frac{dt^{\prime}}{a(t^{\prime})} \, ,
\end{align}
which can be expressed in terms of the Hubble rate $H$ as
\begin{align}
\Delta\eta = \sqrt{-k}\int_{a_i}^{a_0}\frac{da}{a^2 H}  \, .
\end{align}
We introduce the variables $\tilde{a}=a/a_0$ where $a_0$ is the scale factor today, and $\widetilde{H}=H/H_0$ where $H_0$ is the Hubble rate today.  In terms of these variables the Hubble rate is 
\begin{align}
\widetilde{H}(\tilde{a})=\sqrt{\Omega_r\,\tilde{a}^{-4}+\Omega_m \,\tilde{a}^{-3}+\Omega_k\, \tilde{a}^{-2}+\Omega_{\Lambda}}
\end{align}
and so
\begin{align}
f(\tilde{a})=\frac{1}{\tilde{a}^2\widetilde{H}(\tilde{a})}=\frac{1}{\sqrt{\Omega_r+\Omega_m \,\tilde{a}+\Omega_k\, \tilde{a}^{2}+\Omega_{\Lambda}\,\tilde{a}^4}} \, .
\end{align}
The conformal time is then
\begin{align}
\Delta\eta = \frac{\sqrt{-k}}{a_0 H_0}\int_{\tilde{a}_i}^{1} d\tilde{a} f(\tilde{a})  = \sqrt{\Omega_k}\int_{\tilde{a}_i}^{1} d\tilde{a} f(\tilde{a})  \, .
\end{align}
where we have used the definition of the curvature density $\Omega_k={-k}/({a_0^2H_0^2})
$.
We write 
\begin{align}
\zetao=\zetao(\Omega_r,\Omega_m,\Omega_k,\Omega_{\Lambda})=\int_{\tilde{a}_i}^{1} d\tilde{a} f(\tilde{a})
\end{align}
Typically $\tilde{a}_i$ is the scale factor at reheating and we can safely take $\tilde{a}_{i} \simeq 0$.
We set $a_i \rightarrow 0$ in what follows.  To make further progress we split the integral about $\tilde{a}_e=(\Omega_r/\Omega_m)$, the scale factor at matter-radiation equality as
\begin{align}
\zetao \simeq \int_{0}^{\tilde{a}_e} d\tilde{a} f(\tilde{a})+\int_{\tilde{a}_e}^{1} d\tilde{a} f(\tilde{a}
\end{align}
As it starts deep in the radiation dominated epoch and ends at matter-radiation equality the first part of the integral can be very accurately approximated in the limit $\Omega_{\Lambda} \rightarrow 0$ and $\Omega_k \rightarrow 0$ as

\begin{align}
\int_{0}^{\tilde{a}_e} d\tilde{a} f(\tilde{a}) \simeq (\sqrt{2}-1)\frac{2\sqrt{\Omega_r}}{\Omega_m}
\end{align}
The second part of the integral can be solved as a series expansion about $\Omega_k=\Omega_r=0$.
We write
\begin{align}
f(\tilde{a})&=\left. \sum_{m=0}^{\infty}\sum_{n=0}^{\infty}\frac{1}{m! n!}\frac{\partial^{(m+n)} f}{\partial\Omega_k^m \partial\Omega_r^n}\right|_{\stackrel{\scriptstyle \Omega_k=0}{\Omega_r=0}}\Omega_k^m\Omega_r^n \nonumber
\\ &= \left. \sum_{n=0}^{\infty}\frac{1}{n!}\frac{\partial^{n} f}{\partial\Omega_r^n}\right|_{\stackrel{\scriptstyle \Omega_k=0}{\Omega_r=0}}\Omega_r^n +\left.\Omega_k\frac{\partial f}{\partial\Omega_k}\right|_{\stackrel{\scriptstyle \Omega_k=0}{\Omega_r=0}} + \dots
\end{align}
where in second line we have only included the terms that contribute at leading order. We perform the integral term by term to find
\begin{align}
\label{eq:exact}
\int_{\tilde{a}_e}^{1} d\tilde{a} f(\tilde{a})= \sum_{n=0}^{\infty}\frac{(-1)^{n+1}(2n-1)!!}{2^n (4n+1) n!}\frac{1}{\Omega_{\Lambda}^{n+1/2}} \Phi_n +\Omega_k \Psi  + \dots
\end{align}
where
\begin{align}
\Phi_n =\,_{2}F_{1}\left(\fracscr{2n+1}{2}, \fracscr{4n+1}{3}, \fracscr{4n+4}{3};-\fracscr{\Omega_m}{\Omega_{\Lambda}}\right)-\frac{1}{\tilde{a}_e^{4n+1}}\,_{2}F_{1}\left(\fracscr{2n+1}{2}, \fracscr{4n+1}{3}, \fracscr{4n+4}{3};-\fracscr{\Omega_m}{\tilde{a}_e^3\Omega_{\Lambda}}\right)
\end{align}
and
\begin{align}
\Psi=\frac{1}{3\Omega_m^{{3}/{2}}}\left[\left(\frac{1}{\tilde{a}_e^3}+\frac{\Omega_{\Lambda}}{\Omega_m}\right)^{-1/2}-\left(1+\frac{\Omega_{\Lambda}}{\Omega_m}\right)^{-1/2}\right] \, .
\end{align}
Combining these expressions, expanding to leading order in the cosmological parameters, and substituting $\tilde{a}_e \equiv \Omega_r/\Omega_m$ we find that
\begin{align}
\label{eq:approx}
\zetao \simeq\frac{\Gamma[\frac{1}{6}]\Gamma[\frac{4}{3}]}{\sqrt{\pi}\Omega_m^{1/3}\Omega_{\Lambda}^{1/6}}-\frac{1}{\sqrt{\Omega_{\Lambda}}}\left[1-\frac{\Omega_m}{8\Omega_{\Lambda}}-\frac{3\Omega_m^2}{56\Omega_{\Lambda}^2}-\frac{\Omega_k(\Omega_m-2\Omega_{\Lambda})}{6\Omega_m \Omega_{\Lambda}}\right]-\frac{2\sqrt{\Omega_r}}{\Omega_m}
\end{align}
to a very good approximation (higher order terms contribute $\lesssim 0.1\%$ near current cosmological parameters).    Note that while Eq.~(\ref{eq:exact}) is valid for any value of $\Omega_{\Lambda}$ including $\Omega_{\Lambda} \rightarrow 0$, Eq.~({\ref{eq:approx}) is a good approximation only in a Universe, like our own, in which the cosmological constant is already dominating the expansion.

Current cosmological observations \cite{Hinshaw:2008kr}
admit $\Omega_k \sim 0.01$.  For instance, recent cosmological constraints are compatible with $\Omega_{\Lambda} = 0.72 \pm 0.03$ and $\Omega_r = 4.12 \times 10^{-5}/h^2 \simeq (8.3 \pm 0.6) \times 10^{-5}$ near $\Omega_k \simeq 0.005$ or $\Omega_{\Lambda} \simeq 0.71 \pm 0.01$ and $\Omega_r \simeq (7.9 \pm 0.2) \times 10^{-5}$ near $\Omega_k \simeq 0.01$.    These translate to a typical ranges $\zetao \simeq 3.35^{+0.16}_{-0.14}$ near $\Omega_k\simeq 0.005$ and $\zetao \simeq 3.33^{+0.05}_{-0.05}$ near $\Omega_k \simeq 0.01$ --- more than $60\%$ larger than the matter-dominated estimate ($\zeta \simeq 2$) discussed in Section~\ref{sec:conf} of this paper.

\bibliographystyle{utphys}

\begin{thebibliography}{}

\bibitem{Bousso:2000xa}
  R.~Bousso and J.~Polchinski,
  ``Quantization of four-form fluxes and dynamical neutralization of the
  cosmological constant,''
  JHEP {\bf 0006}, 006 (2000)
  [arXiv:hep-th/0004134].

\bibitem{Susskind:2003kw}
  L~Susskind,
  ``The anthropic landscape of string theory,''
  arXiv:hep-th/0302219.





\bibitem{Guth:1982pn}
  A.~H.~Guth and E.~J.~Weinberg,
  ``Could The Universe Have Recovered From A Slow First Order Phase
  Transition?,''
  Nucl.\ Phys.\  B {\bf 212}, 321 (1983).

\bibitem{ggv}
  J.~Garriga, A.~H.~Guth and A.~Vilenkin,
  ``Eternal inflation, bubble collisions, and the persistence of memory,''
  Phys.\ Rev.\  D {\bf 76}, 123512 (2007)
  [arXiv:hep-th/0612242].

\bibitem{aj1}
  A.~Aguirre, M.~C.~Johnson and A.~Shomer,
  ``Towards observable signatures of other bubble universes,''
  Phys.\ Rev.\  D {\bf 76}, 063509 (2007)
  [arXiv:0704.3473 [hep-th]].


\bibitem{aj2}
  A.~Aguirre and M.~C.~Johnson,
  ``Towards observable signatures of other bubble universes II: Exact solutions
  for thin-wall bubble collisions,''
  Phys.\ Rev.\  D {\bf 77}, 123536 (2008)
  [arXiv:0712.3038 [hep-th]].

\bibitem{aj3}
  A.~Aguirre, M.~C.~Johnson and M.~Tysanner,
  ``Surviving the crash: assessing the aftermath of cosmic bubble collisions,''
  arXiv:0811.0866 [hep-th].

\bibitem{bfy}
  R.~Bousso, B.~Freivogel, and I.~Yang, {\it Unpublished}


\bibitem{dahlen}
  A.~Dahlen,
  ``Odds of observing the multiverse,''
  arXiv:0812.0414 [hep-th].


\bibitem{hms}
  S.~W.~Hawking, I.~G.~Moss and J.~M.~Stewart,
  ``Bubble Collisions In The Very Early Universe,''
  Phys.\ Rev.\  D {\bf 26}, 2681 (1982).


\bibitem{fhs}
  B.~Freivogel, G.~T.~Horowitz and S.~Shenker,
  ``Colliding with a crunching bubble,''
  JHEP {\bf 0705}, 090 (2007)
  [arXiv:hep-th/0703146].

\bibitem{ckl1}
  S.~Chang, M.~Kleban and T.~S.~Levi,
  ``When Worlds Collide,''
  JCAP {\bf 0804}, 034 (2008)
  [arXiv:0712.2261 [hep-th]].
  
  
\bibitem{ckl2}
  S.~Chang, M.~Kleban and T.~S.~Levi,
  ``Watching Worlds Collide: Effects on the CMB from Cosmological Bubble
  Collisions,''
  arXiv:0810.5128 [hep-th].






\bibitem{fkrs}
  B.~Freivogel, M.~Kleban, M.~Rodriguez Martinez and L.~Susskind,
  ``Observational consequences of a landscape,''
  JHEP {\bf 0603}, 039 (2006)
  [arXiv:hep-th/0505232].




  
\bibitem{kks}
  N.~Kaloper, M.~Kleban and L.~Sorbo,
  ``Observational implications of cosmological event horizons,''
  Phys.\ Lett.\  B {\bf 600}, 7 (2004)
  [arXiv:astro-ph/0406099].


\bibitem{Loeb:2003ya}
  A.~Loeb and M.~Zaldarriaga,
  ``Measuring the small-scale power spectrum of cosmic density fluctuations
  through 21-cm tomography prior to the epoch of structure formation,''
  Phys.\ Rev.\ Lett.\  {\bf 92}, 211301 (2004)
  [arXiv:astro-ph/0312134].


\bibitem{Kleban:2007jd}
  M.~Kleban, K.~Sigurdson and I.~Swanson,
  ``Cosmic 21-cm Fluctuations as a Probe of Fundamental Physics,''
  JCAP {\bf 0708}, 009 (2007)
  [arXiv:hep-th/0703215].



\bibitem{Mao:2008ug}
  Y.~Mao, M.~Tegmark, M.~McQuinn, M.~Zaldarriaga and O.~Zahn,
 ``How accurately can 21 cm tomography constrain cosmology?,''
  Phys.\ Rev.\  D {\bf 78}, 023529 (2008)
  [arXiv:0802.1710 [astro-ph]].


\bibitem{fk}
  B.~Freivogel and M.~Kleban,
  ``A Conformal Field Theory for Eternal Inflation,''
  to appear.
    
\bibitem{Hinshaw:2008kr}
  G.~Hinshaw {\it et al.}  [WMAP Collaboration],
  Astrophys.\ J.\ Suppl.\  {\bf 180}, 225 (2009)
  [arXiv:0803.0732 [astro-ph]].

    
    
    
\end{thebibliography}

\end{document}